\documentclass[a4paper,12pt,number]{elsarticle}

\usepackage{color}
\usepackage{graphicx}
\usepackage{geometry}
\usepackage{xcolor}
\usepackage[normalsize]{subfigure}
\usepackage{lineno}
\usepackage{array}
\usepackage{ulem}

\usepackage{listings}
\usepackage{mathtools}
\usepackage{leftidx}
\usepackage[bbgreekl]{mathbbol}
\usepackage{amssymb,amsmath}
\usepackage{bbold}
\DeclareSymbolFontAlphabet{\mathbb}{AMSb}
\DeclareSymbolFontAlphabet{\mathbbl}{bbold}

\newcommand{\save}[1]{}

\newcommand{\todo}[1]{}
\newcommand{\rev}[1]{{\color{black}#1}}

\newcommand{\skipToDefault}{
  \thinmuskip=3mu
  \medmuskip=4mu plus 2mu minus 4mu
  \thickmuskip=5mu plus 5mu
}
\newcommand{\domain}{\ensuremath{\Omega}}
\newcommand{\boundary}{\ensuremath{\partial\Omega}}
\newcommand{\C}{\ensuremath{C}}
\newcommand{\stiff}{\ensuremath{\tensf{C}}}

\renewcommand{\S}{\ensuremath{S}}
\newcommand{\eshtens}{\tensf{\S}}
\newcommand{\B}{\ensuremath{\tensf{Q}}}
\renewcommand{\emph}[1]{{\it#1}}
\newcommand{\nill}{\ensuremath{\vek{0}}}

\newcommand{\stiffmicr}{\tensf{\C}}
\newcommand{\complmicr}{\tensf{\C}^{-1}}
\newcommand{\stiffmacr}{\mathbb{\C}}
\newcommand{\complmacr}{\mathbb{\C}^{-1}}
\newcommand{\identity}{\mathbb{I}}

\usepackage{xspace}
\usepackage{amsmath}

\newcommand{\tableset}[2]{\renewcommand{\arraystretch}{#1} \setlength\tabcolsep{#2pt}}

\newcommand{\ctab}[2]{ \multicolumn{1}{#1}{#2} }

\newcommand{\myemph}[1]{\emph{#1}}
\newcommand{\aname}[1]{#1} 
\newcommand{\mname}[1]{\emph{#1}} 

\newcommand{\code}[1]{\tt{#1}}
\newcommand{\mumech}[0]{{$\mu$M\footnotesize{ECH}}}

\newcounter{alglinenumber}
\newcommand{\algstep}{\hspace{5mm}}
\newcommand{\algnum}{\addtocounter{alglinenumber}{1}{\scriptsize \sf \thealglinenumber}}
\newcommand{\algline}[1]{\algnum& #1 \\} 
\newcommand{\function}[1]{{\bf #1}\,}

\newenvironment{algorithm}[1]{%
\setcounter{alglinenumber}{0}
\begin{center}
\begin{tabular}{|rl|}
\hline
& \function{#1} \\
}{ 
\hline
\end{tabular}
\end{center}
\vspace*{-4mm}
}

\newcommand\vek[1]{\mathbf{#1}} 
\newcommand{\tenss}[1]{\bmath{#1}} 
\newcommand{\tensf}[1]{\bmath{\mathbf{#1}}} 
\newcommand{\tensd}[1]{\bmath{\mathcal{#1}}} 

\newcommand{\norm}[1]{\|#1\|}
\newcommand{\measure}[1]{|#1|}

\newcommand{\dcontr}{\,\colon}

\newcommand{\strain}{\varepsilon}
\newcommand{\stress}{\sigma}
\newcommand{\Strain}{E}
\newcommand{\Stress}{\Sigma}

\newcommand{\x}{\vek{x}}

\newcommand{\vfrac}{c}

\newcommand\de[1]{\,{\mathrm d}#1}

\newcommand{\cvfrac}{c_r}
\newcommand{\scompl}[1]{\overline{#1}}

\newcommand{\reffig}  [1]{Fig.~\ref{fig:#1}}

\newcommand{\reftab} [1]{Tab.~\ref{tab:#1}}

\newcommand{\Erefs}[2]{Eqs.~(\ref{#1}--\ref{#2})}

\newcommand{\ErefsIII}[2]{Eqs.~(\ref{#1}) and (\ref{#2})}
\newcommand{\Eref}[1]{Eq.~(\ref{#1})}

\newcommand{\Sref}[1]{Section~\ref{#1}}
\newcommand{\Pref}[1]{Paragraph~\ref{#1}}
\newcommand{\Fref}[1]{Fig.~\ref{#1}}

\usepackage{color}	

\definecolor{orange}{rgb}{1,0.27058824,0}
\definecolor{purple}{rgb}{0.65098039,0.1254902,0.96862745}
\definecolor{violet}{rgb}{0.84313725,0.1254902,0.58823529}

\definecolor{darkblue}{rgb}{0,0,0.55686275}
\definecolor{darkred}{rgb}{0.55686275,0,0}
\definecolor{darkgreen}{rgb}{0,0.39607843,0}
\definecolor{darkcyan}{rgb}{0,0.54117647,0.55686275}
\definecolor{darkmagenta}{rgb}{0.55686275,0,0.55686275}
\definecolor{darkorange}{rgb}{1,0.55686275,0}

\newcommand{\http}[1]{{\tt http://#1}}

\def\SiO2{\rm SiO_2}
\def\Al2O3{\rm Al_2O_3}

\def\Fe2O3{\rm Fe_2O_3}

\def\H2O{\rm H_2O}
\def\Na2O{\rm Na_2O}
\def\SO3{\rm SO_3}
\def\CO2{\rm CO_2}
\def\K2O{\rm K_2O}
\def\P2O5{\rm P_2O_5}
\def\TiO2{\rm TiO_2}
\def\CaCO3{\rm CaCO_3}
\def\CaOH2{\rm Ca \left(OH\right)_2}

\def\CaSO4{\rm CaSO_4}

\def\CaCl2{\rm CaCl_2}

\def\FH3{\rm FH_3}

\newcommand{\bmath}[1]{\mbox{\boldmath$#1$}}

\newcommand{\Tref}[1]{Tab.~\ref{#1}}

\def\imagetop#1{\vtop{\null\hbox{#1}}}

\usepackage[]{hyperref}

\makeatletter
\def\ps@pprintTitle{
	\let\@oddhead\@empty
	\let\@evenhead\@empty
	\def\@oddfoot{{\small\textcopyright 2016. This manuscript version is made available under the \href{http://creativecommons.org/licenses/by-nc-nd/4.0/}{ CC-BY-NC-ND 4.0 license.}}\hfill{}}%
	\let\@evenfoot\@oddfoot
}
\makeatother

\journal{Advances in Engineering Software}

\begin{document}

\begin{frontmatter}

  \title{\mumech{} Micromechanics Library\tnoteref{t1}}
  \tnotetext[t1]{The post-print manuscript of the article published in \mbox{\textit{Advances in Engineering Software}}, \href{http://dx.doi.org/10.1016/j.advengsoft.2016.07.010}{DOI: 10.1016/j.advengsoft.2016.07.010}.}

  \author[ctu]{Ladislav Svoboda}
  \author[ctu]{Stanislav \v{S}ulc}
  \author[ctu]{Tom\'{a}\v{s} Janda}
  \author[ctu]{Jan Vorel}

  \author[ctu]{Jan Nov\'{a}k\corref{cor}}
  \ead{novakj@cml.fsv.cvut.cz}
  \cortext[cor]{Corresponding author. Tel.:~+420-224-354-606}

  \address[ctu]{Faculty of Civil Engineering, Czech Technical University in Prague, Th\'{a}kurova 7, \mbox{166 29 Praha 6}, Czech Republic}

  \begin{abstract}
    The paper presents the project of an open source C/C++ library of analytical solutions to micromechanical fields within media with ellipsoidal heterogeneities. The solutions are based on Eshelby's stress-free, in general polynomial, eigenstrains and equivalent inclusion method. To some extent, the interactions among inclusions in a non-dilute medium are taken into account by means of the self-compatibility algorithm. Moreover, the library is furnished with a powerful I/O interface and conventional homogenization tools. Advantages and limitations of the implemented strategies are addressed through comparisons with reference solutions by means of the Finite Element Method.
  \end{abstract}


\end{frontmatter}


\section{Introduction}\label{sec:intro}
%
In this paper we present a C/C++ library of analytical solutions to classical micromechanical problems. In particular, the library {\mumech} provides users with routines evaluating perturbations of local mechanical fields as strains, stresses, and displacements within a composite medium consisting of isolated ellipsoidal inhomogeneities embedded in an infinite matrix. The implemented, purely analytical solutions to both internal and external fields, i.e. inside and outside inclusion domains, are based on the influential \aname{J. D. Eshelby} work~\cite{eshelby1957determination} and are accomplished in two and three dimensions. The library deals with the heterogeneity problem by means of the equivalent inclusion method. It substitutes heterogeneities with appropriate inclusions subjected to transformation stress free eigenstrains consistent with applied far-field strains~\cite{eshelby1957determination}. Both, constant and polynomial transformation eigenstrains are allowed. The latter is conveniently used to deal with the interacting multiple inclusions. In particular, the interactions among inclusions in a non-dilute media are taken into account by means of the so called self-compatibility algorithm, the fixed version of its ill-posed predecessor reported in~\cite{novak2012micromechanics}. In multiple inclusion problems, contact points among inclusions are allowed however penetrations are not.

\mumech{} was principally designed as a subroutine of Finite Element packages (FEM), justifying so a generic structure of the code and I/O interfaces. It is capable providing Generalized Finite Element environments with subscale enrichment functions to take into account perturbations in macro-field patterns due to microstructural details so as to avoid homogenization based upscaling~\cite{novak2012micromechanics}. Nonetheless, in order to comply with expectations of the micromechanics community, the library has been equipped with several homogenization routines based on direct numerical integration of local fields or conventional techniques as dilute approximation~\cite{eshelby1957determination}, Mori-Tanaka~\cite{Mori1973571}, and Self-consistent schemes~\cite{vsejnoha2013micromechanics}.

The paper is structured as follows. In \Sref{sec:background} we introduce the theoretical background of implemented techniques. In particular, we start with the definition of perturbation fields, give some basics to the Equivalent inclusion method, continue with a brief exposition to the aspects of Eshelby solution due to polynomial stress-free eigenstrains, self-compatibility algorithm and conclude with a summary on homogenization schemes. In \Sref{sec:implementation} we comment on the architecture of \mumech{}, the structure of I/O interfaces, and license regulations. Numerical examples compared with reference solutions by means of FEM are discussed in~\Sref{sec:examples}. Final remarks concluding the exposition are given in~\Sref{sec:conclusions}.

\section{Background}\label{sec:background}
%
In what follows, we give a very brief introduction to theoretical background of implemented strategies in \mumech{} library. The entire \Sref{sec:background} can be omitted by readers versed in classical micromechanics. As for the notation used throughout the section, we mostly use the compact tensorial form denoted by different font styles in bold depending on particular order of the tensors. However, where the exposition requires, we resort to standard tensorial notation with indices. For instance,
\begin{align*}
  c_i = A_{ijkl}a_{kl}b_j = \tensf{A}\dcontr\tenss{a}\cdot\vek{b} = \vek{c}
\end{align*}
Also, note that the superscripts over state variables do not stand for power indices. In the case of stiffness or concentration tensors, in general fourth order tensors, analogical indices are written as subscripts and superscripts are reserved e.g. for $\bullet^{-1}$ inverse operator. In addition, $i,j,k,l,m,q$ are reserved for tensorial indices, $r,s$ denote the inclusion enumerators, and $p$ stands for the iteration loop increment.

\subsection{Perturbation fields}\label{ss:pert_fiedlds}
\begin{figure}[!ht]
  \centering
  \includegraphics[width=45mm]{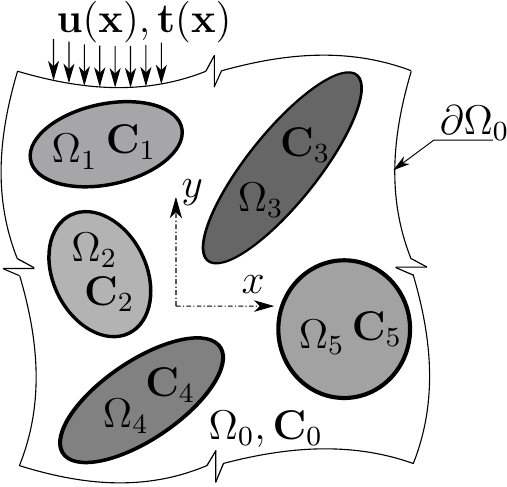}
  \caption{Infinite composite body with ellipsoidal inhomogeneities.}
  \label{fig:approx_fces_solu_strategy}
\end{figure}

%
Let assume an infinite isotropic homogeneous body with separated heterogeneities, \Fref{fig:approx_fces_solu_strategy}. Moreover, consider the body be the subject to some macroscopic excitation, e.g. a remote strain induced by a combination of displacement and traction fields $\vek{u}(\x), \vek{t}(\x)$ applied to the boundary $\boundary_0$ at infinity.
The stiffness of such a composite can be decomposed as~\cite{mura1987micromechanics,Novak:2008:CESS}
\begin{equation}
  \stiff(\x) = \stiff_0 + \overline{\stiff}(\x)
  \label{eq:stiff_decomposition}
\end{equation}
where $\stiff_0$ represents the fourth order tensor of elastic constants of the matrix $\domain_0$ and $\overline{\stiff}(\x) = \stiff(\x) - \stiff_0$ is its piecewise constant complement to $\stiff(\x)$ due to the presence of $1,\dots,n$ heterogeneities. Note, $\overline{\stiff}(\x)$ is nonzero only for $\x \in \bigcup_{r=1}^n\domain_r$.
As a result of applied loads, the heterogeneous body experiences local fields that can be decomposed by analogy to \Eref{eq:stiff_decomposition} as
\begin{eqnarray}
  \vek{u}(\x) = \vek{u}^0(\x) + \overline{\vek{u}}(\x),&
  \tenss{\strain}(\x) = \tenss{\strain}^0 + \overline{\tenss{\strain}}(\x),&
  \tenss{\stress}(\x) = \tenss{\stress}^0 + \overline{\tenss{\stress}}(\x)
  \label{eq:fields_decomposition}
 \end{eqnarray}
Here, the superscript $\bullet^0$ indicates the homogeneous (macroscopic) component of the state variables in the absence of heterogeneities and $\overline{\bullet}$ stands for its perturbation (microscopic) counterpart induced by their presence.
Determination of the perturbation fields is based on the equivalent inclusion method as proposed by Eshelby in~\cite{eshelby1957determination}. Here, we first limit the exposition to a single ellipsoidal heterogeneity embedded in a homogeneous matrix undergoing a uniform remote strain excitation and then explore some possibilities to take into account interactions among multiple heterogeneities.

\subsection{Equivalent inclusion method for single heterogeneity problem}

\begin{figure}[ht]
  \centering
  \begin{tabular}{>{\centering}m{40mm}>{\centering}m{0mm}>{\centering}m{40mm}>{\centering}m{0mm}%
      >{\centering\arraybackslash}m{40mm}}
    \includegraphics[width=45mm]{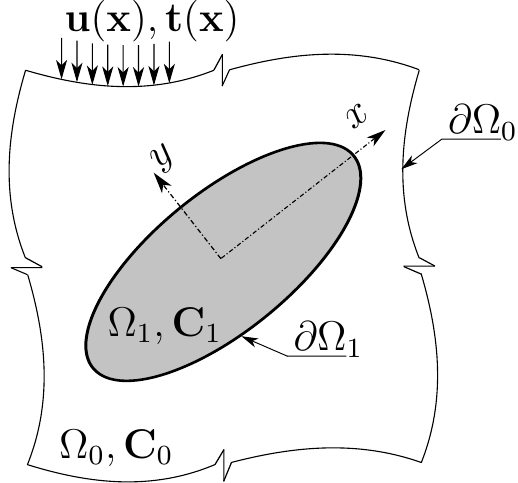} &$\equiv$
    &\includegraphics[width=45mm]{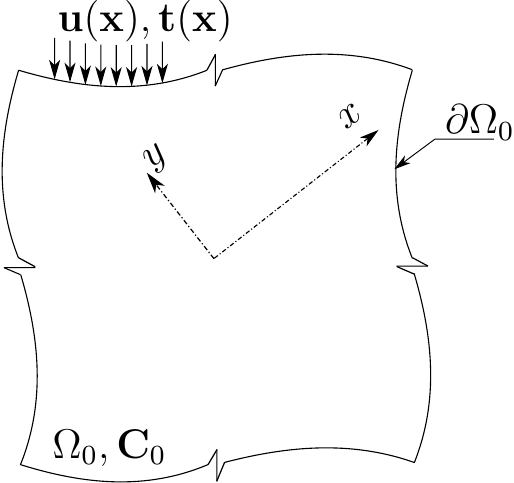}
    &$+$ &\includegraphics[width=45mm]{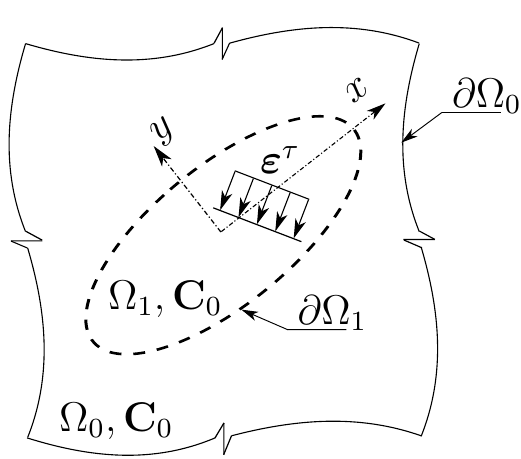}\\
    (a) & &(b) & &(c)
  \end{tabular}
  \caption{Equivalent inclusion method, a) inhomogeneity problem, b) infinite homogeneous body with allied loads, c) homogeneous inclusion problem.}
  \label{fig:eq_incl_method_principle}
\end{figure}

When seeking for local fields by means of the equivalent inclusion method, we replace the heterogeneity problem, \Fref{fig:eq_incl_method_principle}a, by an equivalent inclusion problem consisting of the homogeneous matrix exposed to a suitable stress-free eigenstrain $\tenss{\strain}^{\tau}(\x)$ which vanishes everywhere except for $\x\in\domain_1$, \Fref{fig:eq_incl_method_principle}c, supplement to external loads applied at infinity, \Fref{fig:eq_incl_method_principle}b, see~\cite{eshelby1957determination} for further details.
The solution of the inclusion problem then primarily requires the determination of the transformation eigenstrain $\tenss{\strain}^{\tau}(\x)$ that induces identical perturbation to the homogeneous fields as it would occur due to the original inhomogeneity. As there are no other inclusions surrounding that of our concern, $\tenss{\strain}^{\tau}(\x)$ remains constant in $\domain_1$, that is we can write $\tenss{\strain}^{\tau}(\x\in\domain_1) = \tenss{\strain}^{\tau}$. According to Hook's law and decompositions in \Eref{eq:fields_decomposition}$^{2,3}$, local stresses rendered by the inhomogeneity problem, \Fref{fig:eq_incl_method_principle}a, read as
\begin{equation}
  \tenss{\stress}(\x) = \tenss{\stress}^0 + \overline{\tenss{\stress}}(\x) = \stiff(\x) \dcontr[ \tenss{\strain}^0 + \overline{\tenss{\strain}}(\x)]
  \label{eq:stress_het}
\end{equation}
For the equivalent problem holds
\begin{equation}
  \tenss{\stress}(\x) = \stiff_0 \dcontr [\tenss{\strain}^0 + \overline{\tenss{\strain}}(\x) - \tenss{\strain}^\tau(\x)]
  \label{eq:stress_hom}
\end{equation}
Given the fact that $\tenss{\stress}^0 = \stiff_0\dcontr \tenss{\strain}^0$, it yields from \Eref{eq:fields_decomposition}$^3$ and \Eref{eq:stress_hom}
\begin{equation}
  \overline{\tenss{\stress}}(\x) = \stiff_0 \dcontr [\overline{\tenss{\strain}}(\x) - \tenss{\strain}^\tau(\x)]
  \label{eq:stress_per}
\end{equation}
Now, equating the rhs's of \Erefs{eq:stress_het}{eq:stress_hom},
\begin{equation}
  \stiff(\x) \dcontr [\tenss{\strain}^0 + \overline{\tenss{\strain}}(\x)] = \stiff_0 \dcontr [\tenss{\strain}^0 + \overline{\tenss{\strain}}(\x) - \tenss{\strain}^\tau(\x)]
  \label{eq:stress_equality}
\end{equation}
and taking into account the following fundamental solution for $\overline{\tenss{\strain}}(\x)$
\begin{equation}
  \overline{\tenss{\strain}}(\x) = \eshtens(\x) \dcontr \tenss{\strain}^\tau
  \label{eq:eshelby}
\end{equation}
where $\eshtens(\x)$ denotes the Eshelby tensor evaluated at an arbitrary point $\x$, results
\begin{equation}
  \left[\stiff(\x) - \stiff_0 \right] \dcontr \tenss{\strain}^0 =
  \left[\stiff_0 \dcontr \eshtens(\x) - \stiff(\x) \dcontr \eshtens(\x) - \stiff_0 \right] \dcontr \tenss{\strain}^\tau
  \label{eq:eq_incl_meth_step_4}
\end{equation}
The definition of $\eshtens(\x)$ tensor is as in \Eref{eq:eshtensor_ext} and \Eref{eq:eshtensor_int}$^1$ while the detailed derivation can be found e.g. in~\cite{eshelby1957determination,mura1987micromechanics}.
Finally, \Eref{eq:eq_incl_meth_step_4} gives rise the sought stress free transformation eigenstrain $\tenss{\strain}^\tau$ in the form
\begin{equation}
  \tenss{\strain}^\tau = \B \dcontr \tenss{\strain}^0
  \label{eq:eq_incl_meth_step_5}
\end{equation}
where tensor $\B$ reads as
\begin{equation}
  \B = -\left[\,\overline{\stiff}_1 \dcontr \eshtens(\nill) + \stiff_0 \right]^{-1} \dcontr \overline{\stiff}_1
  \label{eq:remote_transf_strain_oper}
\end{equation}
Once the transformation eigenstrain has been determined, \Eref{eq:eq_incl_meth_step_5}, the stress perturbation can be computed from \Eref{eq:stress_per} and displacement perturbations as
\begin{equation}
  \overline{\vek{u}}(\x) = \tensd{L}(\x)\dcontr\tenss{\strain}^\tau =
  \tensd{L}(\x)\dcontr\B\dcontr\tenss{\strain}^0
  \label{eq:eshelby_disp}
\end{equation}
where the operator $\tensd{L}(\x)$ is the third order Eshelby tensor-like operator mapping $\tenss{\strain}^\tau \mapsto \overline{\vek{u}}(\x)$ whose detailed derivation can be found in~\cite{novak2012micromechanics}.

\subsection{Single inclusion problem for polynomial eigenstrains}\label{sec:polynomials}
%
The solution to a single inclusion in the infinite matrix loaded by a constant eigenstrain stated formally in equation~\Eref{eq:eshelby} was generalized by Sendeckyj (1967) and Moschovidis (1975) for eigenstrains prescribed in a general polynomial form~\cite[and references therein]{mura1987micromechanics}. For a simpler exposition, here we summarize only the solution to the single inclusion under linear eigenstrain excitation, though the solution to the problem with quadratic eigenstrains is also implemented to some extent in current version of \mumech{}. Going back to index based \rev{Einstein summation convention}, the prescribed linear eigenstrain field that is nonzero only for $\x \in\domain_1$, is defined as
\begin{equation}
  \strain^\tau_{ij}(\x) = \strain^\tau_{ij} + \strain^\tau_{ijk} x_k
  \label{eq:linear_eigenstrain}
\end{equation}
where $\strain^\tau_{ij}$ is the constant part of the imposed eigenstrain identical to that from the previous paragraph, and $\strain^\tau_{ijk}$ contains its gradient complements in $k$-th coordinate direction. By analogy to~\Eref{eq:eshelby}, it holds
\begin{equation}
  \overline{\strain}_{ij}(\x) = \S_{ijkl}(\x)\strain^\tau_{kl} + \S_{ijklm}(\x)\strain^\tau_{klm}
  \label{eq:linear_pert_strain}
\end{equation}
The solution thus reduces to seeking for components of tensors $\S_{ijkl}$ and $\S_{ijklm}$, which depend only on the proportions of the semi-axes of $\domain_1$ and matrix Poisson's ratio $\nu_0$. In particular, for exterior points, i.e. $\x \notin \domain_1$, the following definitions hold~\cite{mura1987micromechanics}
\begin{align}
  \S_{ijkl}(\x) &= \rev{[}\psi_{,klij}-2\nu_0\delta_{kl}\phi_{,ij}\nonumber \\
    &-(1-\nu_0)(\phi_{,kj}\delta_{il}+\phi_{,ki}\delta_{jl}+\phi_{,lj}\delta_{ik}+\phi_{,li}\delta_{jk})] /[8\pi(1-\nu_0)]
  \label{eq:eshtensor_ext}
\end{align}
%
\begin{align}
  \S_{ijklm}(\x) &= [\psi_{m,klij}-2\nu_0\delta_{kl}\phi_{m,ij} \nonumber \\
    &- (1-\nu_0)(\phi_{m,kj}\delta_{il}+\phi_{m,ki}\delta_{jl}+\phi_{m,lj}\delta_{ik}+\phi_{m,li}\delta_{jk})] / [8\pi(1-\nu_\rev{0})]
  \label{eq:eshtensor_lin_ext}
\end{align}
%
On the other hand, for all points $\x$ inside $\domain_1$, the above tensors read as
\begin{equation}
  \S_{ijkl}(\x) = \S_{ijkl}(\nill), \quad   \S_{ijklm}(\x) = \S_{ijklm,q}(\nill)x_q
  \label{eq:eshtensor_int}
\end{equation}
Symbol $\delta_{ij}$ denotes the Kronecker delta and $\bullet_{,i}$ stands for the first derivative in $i$-th direction. The potentials $\phi$ and $\phi_i$ and the first derivative of potential $\psi$ and $\psi_i$ are defined as
\begin{align}
  \phi(\x) &= V(\x),\quad
  \phi_i(\x) = a^2_i x_i V_i(\x),\quad
  \psi_{,i}(\x) = x_{i}[V(\x) - a_i^2 V_i(\x)]\nonumber\\
  \psi_{i,j}(\x) &= -\frac{1}{4} \delta_{ij} a^{2}_{i}
  \{V(\x)-x^{2}_{k} V_{k}(\x)-a^{2}_{i} [V_{i}(\x)-x^{2}_{k} V_{ki}(\x)]\}\nonumber\\
  &+ a^{2}_{i} x_{i} x_{j}[V_{j}(\x)-a^{2}_{i}V_{ji}(\x)]
\end{align}
with $a_i$ being ellipsoidal semi-axe lengths sorted in descending order \rev{and index $k$ being the summation index}. Elliptic integrals $V$, $V_i$ and $V_{ij}$, respectively, read
\thinmuskip=0mu
\medmuskip=1mu
\thickmuskip=1mu
\begin{equation}
  V(\x) = \frac{I(\lambda) - x^{2}_{k} I_{k}(\lambda)}{2},\quad
  V_{i}(\x) = \frac{I_{i}(\lambda) - x^{2}_{k} I_{ik}(\lambda)}{2},\quad
  V_{ij}(\x) = \frac{I_{ij}(\lambda) - x^{2}_{k} I_{ijk}(\lambda)}{2}
\end{equation}
\skipToDefault
where $I$, $I_{i}$, $I_{ij}$ and $I_{ijk}$ are $\lambda$-variable dependent elliptic integrals. The value of $\lambda$ for a given point $\x\in\domain_1$ is the largest positive root of the cubic equation
\begin{equation}
  \frac{x_1^2}{a_1^2 + \lambda} + \frac{x_2^2}{a_2^2 + \lambda} + \frac{x_3^2}{a_3^2 + \lambda} = 1
\end{equation}
and zero otherwise. The elliptic integrals $I$ and $I_i$ are expressed as
\medmuskip=1mu
\begin{align}
  I(\lambda) &= b(a^2_1-a^2_3)^{-\frac{1}{2}}F(\theta, c),\nonumber\\
  I_{1}(\lambda) &= b(a^2_1-a^2_2)^{-1}(a^2_1-a^2_3)^{-\frac{1}{2}}[F(\theta, c)-E(\theta, c)],\nonumber\\
  I_{2}(\lambda) &= b[(a^2_1-a^2_2)^{-1}(a^2_2-a^2_3)^{-1}(a^2_1-a^2_3)^{\frac{1}{2}}E(\theta, c)
    - (a^2_1-a^2_2)^{-1}(a^2_1-a^2_3)^{-\frac{1}{2}}F(\theta, c) \nonumber \\
    &- (a^2_2-a^2_3)^{-1}(a^2_3+\lambda)^{\frac{1}{2}}(a^2_1+\lambda)^{-\frac{1}{2}}(a^2_2+\lambda)^{-\frac{1}{2}}], \nonumber\\
  I_{3}(\lambda) &= b(a^2_2-a^2_3)^{-1}(a^2_1-a^2_3)^{-\frac{1}{2}}
  [(a^2_1-a^2_3)^{\frac{1}{2}}(a^2_2+\lambda)^{\frac{1}{2}}(a^2_1+\lambda)^{-\frac{1}{2}}(a^2_3+\lambda)^{-\frac{1}{2}}]
\end{align}
\skipToDefault
with $b=4\pi a_1a_2a_3$, $\theta = \arcsin\sqrt{1-a_3^2/a_1^2}$, and $c=\sqrt{(a_1^2-a_2^2)/(a_1^2-a_3^2)}$. Functions $F$ and $E$ are the incomplete Legendre elliptic integrals defined as
\begin{equation}
  F(\theta, c) = \int_0^\theta{\frac{\de{w}}{\sqrt{1-c^2\sin^2w}}},\quad
  E(\theta, c) = \int_0^\theta{\sqrt{1-c^2\sin^2w}\,\de{w}}
\end{equation}
In addition, higher order integrals $I_{ij}$ and $I_{ijk}$ are expressed by means of those of the lower orders and by substituting $\alpha=(a_i^2-a_j^2)$ as follows
%
\begin{align}
  I_{ij}(\lambda) &= [I_j(\lambda)-I_i(\lambda)]/\alpha,\quad I_{iij}(\lambda) = [I_{ij}(\lambda) - I_{ii}(\lambda)]/\alpha\quad \forall \, i \neq j, \nonumber\\
  I_{ii}(\lambda) &= \frac{1}{3}[\frac{b}{(a_i^2+\lambda)^2\Delta(\lambda)} - I_{ij}(\lambda) - I_{ik}(\lambda)],\quad
  I_{ijk}(\lambda) = [I_{jk}(\lambda) - I_{ik}(\lambda)]/\alpha,\nonumber\\
  I_{iii}(\lambda) &= \frac{1}{5}[\frac{b}{(a_i^2+\lambda)^2\Delta(\lambda)} - I_{iij}(\lambda) - I_{iik}(\lambda)] \quad \forall \, i \neq j \neq k \neq i
\end{align}
%
Finally, $\Delta(\lambda)$ reads as
\begin{equation}
  \Delta(\lambda) = \sqrt{(a_1^2+\lambda)(a_2^2+\lambda)(a_3^2+\lambda)}
\end{equation}

\subsection{Multiple-inclusion problem}\label{sec:selfbalancing}

\begin{figure}[!ht]
  \centering
  \begin{tabular}{cc}
  \includegraphics[height=40mm]{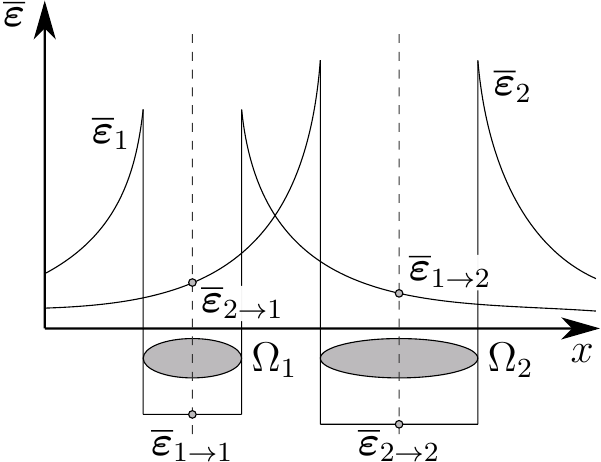} &  \includegraphics[height=40mm]{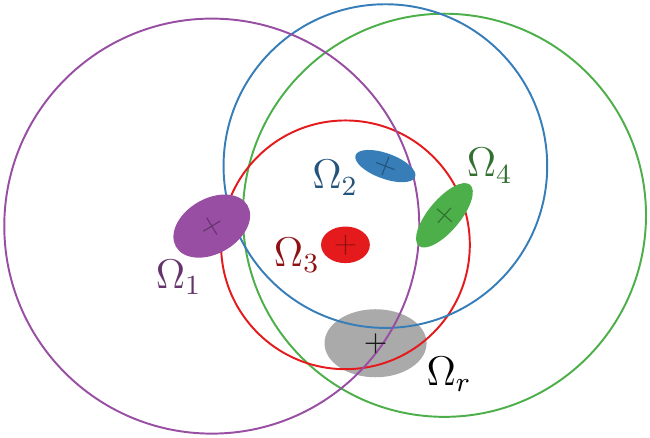}\\
  (a) &(b)
  \end{tabular}
  \caption{Illustration of self-compatibility algorithm, a) principle for double inclusion problem, b) inclusions having non-negligible impact on $r$-th inclusion $\xi_r = \norm{\domain_1, \domain_3, \domain_4} = 3$.}
  \label{fig:self-bal}
\end{figure}

In the case of an infinite matrix with multiple inclusions, the perturbation fields within $\domain_r$, \Fref{fig:approx_fces_solu_strategy}, are no longer uniformly distributed as a result of their mutual interactions. In \mumech{}, we account for the interactions only approximately by assuming the eigenfields within $r$-th inclusion be still constant, however, influenced by local changes of state variables due to the remaining inclusions, namely those nearby $\domain_r$. In particular, we control the ``compatibility'' of the perturbation strain field inside each inclusion calculated by means of \Eref{eq:eshelby}. The key ingredient of these formulas, $\tenss{\strain}^0_r$, mapped to $\tenss{\strain}^\tau_r$ through $\B_r$, is recursively increased by perturbation strains $\overline{\tenss{\strain}}_{s\to r}$ arising from the presence of $s=1,\dots,n$ inclusions. That is why, we have different $\tenss{\strain}^0_r$ for each of $1,\dots,r,\dots,n$ inclusions. Individual contributions $\overline{\tenss{\strain}}_{s\to r}$ to $\tenss{\strain}^{0,\mathrm{tot}}_r$ are measured in the center of $\domain_r$, \Fref{fig:self-bal}a. Thus, $\tenss{\strain}^{0,\mathrm{tot}}_r$ in the $p$-th iteration loop of the self-compatibility procedure reads as
\begin{equation}
  \leftidx{^p}{\tenss{\strain}}{^{0,\mathrm{tot}}_r} \stackrel{\mathrm{def}}{=} \tenss{\strain}^0 + \sum_{s\backslash r}^{n} \leftidx{^p}{\overline{\tenss{\strain}}}_{s\to r}
  \label{eq:selfcompatibility_1}
\end{equation}
where the contributions $\leftidx{^p}{\overline{\tenss{\strain}}}_{s\to r}$ are evaluated from the previous remote field $\leftidx{^{p-1}}{\tenss{\strain}}{^{0,\mathrm{tot}}_r}$\rev{; the ${s\backslash r}$ operation excludes inclusion $r$ from the set of $n$ inclusions.} The initial remote strain $\tenss{\strain}^0$ is imposed to the matrix surrounding all inclusions at the beginning of the procedure, i.e. $\leftidx{^1}{\tenss{\strain}}{^{0,\mathrm{tot}}_r} = \tenss{\strain}^0$. The line-by-line definition of the iterative algorithm based on \Eref{eq:selfcompatibility_1} follows in \Tref{tab:selfBalancing}.
\begin{table}[!ht]
  \begin{algorithm}{Self Compatibility Algorithm $(\tenss{\strain}^0_r, \tenss{\strain}^\tau_r, \B_r, \tensf{S}_r, n)$}
    \algline{\function{Do}}
    \algline{\algstep \function{For} $(r \leq n)$}
    \algline{\algstep \algstep
      $\tenss{\strain}^{\tau,\mathrm{prev}}_{r} = \tenss{\strain}^\tau_{r}$}
    \algline{\algstep \algstep $\tenss{\strain}^{0,\mathrm{tot}}_r = \tenss{\strain}^{0} + \sum_{s\backslash r}^n\overline{\tenss{\strain}}_{s\to r}$}
    \algline{\algstep \algstep $\tenss{\strain}^\tau_r = \tensf{B}_r\dcontr\tenss{\strain}^{0,\mathrm{tot}}_r$}
    \algline{\algstep \algstep
      $\Delta\tenss{\strain}^\tau_r =  \tenss{\strain}^\tau_{r} - \tenss{\strain}^{\tau,\mathrm{prev}}_{r}$}
    \algline{\algstep \function{EndFor}}
    \algline{\function{While} $\big(\sum_{r}^n\norm{\Delta\tenss{\strain}^\tau_{r}} > \eta \big)$}
  \end{algorithm}
  \caption{\mname{Self-compatibility} algorithm. In principle, within each iteration loop we consider the effect of $s$-th inclusion on inclusion $r$ as an additional external load entering the solution to its equivalent stress-free eigenstrain $\tenss{\strain}^\tau_r$. This means, in each iteration, we recalculate $\tenss{\strain}_r^{0,tot}$ for every single inclusion as the sum of the prescribed homogeneous strain $\tenss{\strain}^0$ and its perturbations due to remaining inclusions $\overline{\tenss{\strain}}_{s\to r}$ evaluated in the centroid of inclusion $r$, line 4. Next, $\tenss{\strain}_r^{\tau}$ is updated, line 5. The algorithm continues until an acceptable tolerance $\eta$ between the Euclidean norms of the two consecutive stress-free eigenstrains $\strain^\tau_r$ is achieved, line 8.}
  \label{tab:selfBalancing}
\end{table}
At its convergence, the stress and displacement perturbations corresponding to compatible transformation eigenstrains are recalculated according to \Eref{eq:stress_per} and \Eref{eq:eshelby_disp}. It is worthwhile to note that the algorithm does not depend on a particular sequence of inclusions, as follows from the elastic reciprocity theorem~\cite[and references therein]{pichler2010estimation}. The iterative procedure has been chosen since a closed form solution for the multiple inclusion problem does not exist and a numerical one would be prohibitively expensive, see e.g.~\cite{pichler2010estimation}. The computational complexity of the so called \emph{full} version of the algorithm is $O(n^2)$. However, this can be further reduced by taking into account only inclusions that have a non-negligible impact on the $r$-th inclusion of interest, usually those placed very nearby $\domain_r$ or excessively large inclusions in the case of somehow disparate polydisperse. This algorithm is called \emph{optimized} in \mumech{}. Its complexity reduces to $O(\xi n)$, where $\xi=\frac{1}{r}\sum_r\xi_r$ is the arithmetic average of the number of inclusions whose cut outs limited by radii\footnote{A usual choice is $2.5$ multiple of the longest semi-axis $a_1$.} $\rho_s$ centered in $\x_s$ embrace the $r$-th inclusion, \Fref{fig:self-bal}b. Note, for $n\to\infty$ the complexity of \emph{optimized} algorithm is $O(n)$ as $\xi=\mathrm{const}\ll n$.

\subsection{Approximation to perturbation strain concentrations}\label{sec:approx_concentrations}

When dealing with a dilute distribution of inclusions, constant strain and stress fields are assumed within $\domain_r$ and no question on concentrations comes in play. However, the goal of \mumech{} is to proceed a few steps beyond, namely to non-dilute dispersions. The concentrations arising from mutual inclusion interactions are approximated by the following procedure, which stems from the approximate solution to the sought non-constant transformation eigenstrain $\tenss{\strain}_r^{\tau}(\x)$ suitable for the decomposition in \Eref{eq:linear_eigenstrain} or similar one of a higher degree.

Consider point a $\x$ inside the inclusion domain $\domain_r$, where we calculate the stress free transformation eigenstrain as
\begin{equation}
  \tenss{\strain}_r^{\tau}(\x) \approx \B_r\dcontr \bigg[ \tenss{\strain}^0 + \sum_{s\backslash r}^n\overline{\tenss{\strain}}_{s\to r}(\x)\bigg]
  \label{eq:post_process_I}
\end{equation}
Consequently we sample $\tenss{\strain}_r^{\tau}(\x)$ in as many points $\x$ as necessary for a polynomial of chosen degree, three in the case case of \Eref{eq:linear_eigenstrain}. Finally, the solution to perturbation fields for points $\x$ in each domain $\domain_r$ are obtained following the exposition given in \Pref{sec:polynomials}. For the points within the matrix, i.e. $\x$ outside the union $\bigcup_{r=1}^n\domain_r$, the solutions are obtained by the sum of individual contributions attributed to each of $1,\dots,n$ inclusions.

An alternative, and surprisingly well working, approach such that it does not call for the implementation of polynomial eigenstrain problem is, that the solutions to perturbation strains in the internal points are calculated by means of the basic Eshelby formula given in \Eref{eq:eshelby}, as
\begin{equation}
  \overline{\tenss{\strain}}_r(\x) \approx \eshtens_r \dcontr \tenss{\strain}_r^{\tau}(\x) + \sum_{s\backslash r}^n\overline{\tenss{\strain}}_{s\to r}(\x)
\end{equation}
where $\tenss{\strain}_r^{\tau}(\x)$ is that provided by \Eref{eq:post_process_I}. By analogy to the latter approach, the solution in external points is obtained by adding up contributions from all inclusions.

Finally, it is worthwhile to note that the computational complexity can be controlled by the number of internal points chosen to approximate $\tenss{\strain}_r^{\tau}(\x)$ by either of the approaches above. In addition, substantial savings can be made by choosing \myemph{optimized} mode running on the same principles as in the case of the self-compatibility algorithm.

\subsection{Homogenization}\label{sec:homogenization}

Assuming non-elastic phenomena be entirely attributed to the microstructure evolution dynamics, the constitutive behavior of an arbitrary point $\vek{x}$ at an instant is governed by the following pair of equations~\cite{zeman2003analysis},
\begin{equation}
  \tenss{\stress}_r(\vek{x}) = \stiffmicr_r\dcontr\tenss{\strain}_r(\vek{x}),\quad
  \tenss{\strain}_r(\vek{x}) = \complmicr_r\dcontr\tenss{\stress}_r(\vek{x})\quad \mathrm{for}\quad \vek{x}\in \domain_r
\end{equation}
where $r=0,\dots,n$. According to Hill's lemma~\cite{Hill1963357,zeman2003analysis}, the averages of the above local quantities $\tenss{\stress}(\x)$ and $\tenss{\strain}(\x)$ are coupled with their macroscopic conjugates $\tenss{\Stress}, \tenss{\Strain}$ as
\thinmuskip=0mu
\medmuskip=1mu
\begin{eqnarray}
  \langle\tenss{\stress}(\vek{x})\rangle = \langle\stiffmicr(\vek{x})\dcontr\tenss{\strain}(\vek{x})\rangle = \sum_r^n \cvfrac \stiffmicr_r\dcontr\langle\tenss{\strain}_r(\vek{x})\rangle = \sum_r^n \cvfrac \stiffmicr_r\dcontr\tenss{\strain}_r = \stiffmacr\dcontr\tenss{\Strain}\label{eq:basic_averaging_rule_stiff},\nonumber\\
  \langle\tenss{\strain}(\vek{x})\rangle = \langle\complmicr(\vek{x})\dcontr\tenss{\stress}(\vek{x})\rangle = \sum_r^n \cvfrac \complmicr_r\dcontr\langle\tenss{\stress}_r(\vek{x})\rangle = \sum_r^n \cvfrac \complmicr_r\dcontr\tenss{\stress}_r = \complmacr\dcontr\tenss{\Stress}
\label{eq:basic_averaging_rule_compl}
\end{eqnarray}
\skipToDefault
It is evident, that the effective elastic stiffness and compliance tensors depend on the elastic properties of each phase $\stiffmicr_r$ and volume fractions $\cvfrac$. In addition, they depend on mutual interactions given by the intrinsic geometrical arrangement of the phases and the compatibility or equilibrium requirements, encoded in concentration factors $\tensf{A}_r, \tensf{B}_r$ for which it holds~\cite{walpole1969overall,eshelby1957determination}
\begin{equation}
  \tenss{\strain}_r = \tensf{A}_r\dcontr\tenss{\Strain},\, \tenss{\stress}_r = \tensf{B}_r\dcontr\tenss{\Stress}
  \label{eq:eps_r}
\end{equation}
Plugging the latter definitions in last two terms of \Eref{eq:basic_averaging_rule_compl} gives
\begin{equation}
  \stiffmacr = \sum_r^n \cvfrac \stiffmicr_r\dcontr\tensf{A}_r,\quad
  \complmacr = \sum_r^n \cvfrac \complmicr_r\dcontr\tensf{B}_r
\end{equation}
Now, identifying by $r=0$ a matrix phase in which the remaining heterogeneities are fully embedded, and taking into account the fact that $\vfrac_0 \tensf{A}_0 = \identity - \sum_{r=1}^n \cvfrac\tensf{A}_r$, $\vfrac_0 \tensf{B}_0 = \identity - \sum_{r=1}^n \cvfrac\tensf{B}_r$, where $\identity$ is the fourth order identity tensor, and considering $\vfrac_0 = 1 - \sum_{r=1}^n \cvfrac$, yields
\begin{equation}
  \stiffmacr = \stiffmicr_0 + \sum_r^n \cvfrac \scompl{\stiffmicr}_r\dcontr\tensf{A}_r,\quad
  \complmacr = \complmicr_0 + \sum_r^n \cvfrac \scompl{\stiffmicr}^{-1}_r\dcontr\tensf{B}_r
  \label{eq:C^eff_C^-1,eff}
\end{equation}
where, according to \Eref{eq:stiff_decomposition}, it holds
\begin{equation}
  \scompl{\stiffmicr}_r = \stiffmicr_r - \stiffmicr_0
  \label{eq:\cvfrac_supplements}
\end{equation}
From now on, the central question is how to evaluate the concentration factors $\tensf{A}_r$ and $\tensf{B}_r$ for a medium with multiple inclusions.

\subsubsection{Homogenization by direct integration of approximate local fields}\label{sec:DIM}
%
\begin{figure}[!ht]
  \centering
  \includegraphics[width=45mm]{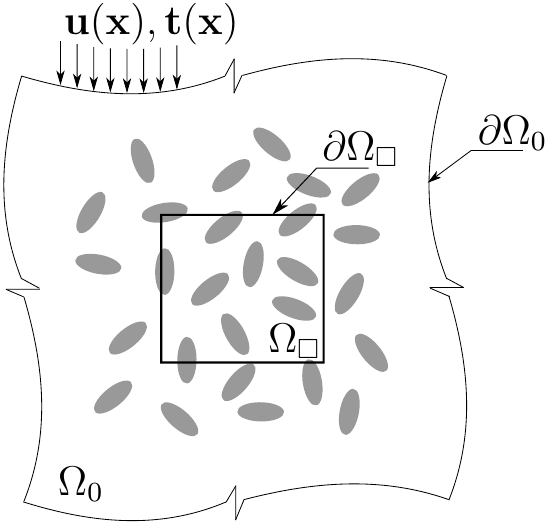}
  \caption{Two-dimensional illustration of DIM subdomain $\domain_\Box$ in cluster of inclusions embedded in infinite matrix $\domain_0$.}
  \label{fig:DIM_subdomain}
\end{figure}

The Direct Integration Method (DIM) stems of the numerical integration of local stresses and strains in the subregion $\domain_\Box$, \Fref{fig:DIM_subdomain}, of a larger cluster of inclusions embedded in the matrix  and arising from the successive load steps by a single unitary component of $\strain_{ij}^0$ while the other vanish, see e.g.~\cite{vsejnoha2013micromechanics}. Thus, the set of nine $ij$-th components\footnote{one column or row in Voight-Mandel notation} of the fourth order tensor of effective stiffness moduli is rendered as
\begin{equation}
  \tensf{\stiffmacr} = \mathbb{E}^{-1} \dcontr \mathbbl{\Stress}
  \label{eq:direct_integration}
\end{equation}
where
\begin{equation}
  \mathbbl{\Stress}_{ijkl} \stackrel{\strain^0_{ij} = 1}{=} \langle\stress_{kl}\rangle = \frac{1}{\measure{\domain_\Box}}\int_{\domain_\Box} \stress_{kl}\de{\domain},\quad
  \mathbb{E}_{ijkl} \stackrel{\strain^0_{ij} = 1}{=} \langle\strain_{kl}\rangle = \frac{1}{\measure{\domain_\Box}}\int_{\domain_\Box} \strain_{kl}\de{\domain}
  \label{eq:direct_integration_2}
\end{equation}
This homogenization procedure assumes the subregion boundary $\partial\domain_\Box$ be sufficiently far from the boundary of the cluster of all inclusions entering the analysis in order to guarantee vanishing boundary effects. In addition, it is considered that the subregion's volume and geometry is representative to the solved microstructure. In other words, it should form its Representative Volume Element (RVE)~\cite{Hill1963357}. It is also worthwhile to note that the shape of $\domain_\Box$ is completely arbitrary. It does not even need to form a continuous domain.

\subsubsection{Dilute approximation}

Suppose the dispersion of inclusions distributed in the infinite matrix is low or, say, dilute. Under such conditions, inclusions do not interact, and as a consequence, the macroscopic strain $\tenss{\Strain}$ from \Eref{eq:basic_averaging_rule_compl} and \Eref{eq:eps_r} can be imagined as equal to the remote strain $\tenss{\strain}^0$ from the exposition introduced in \Pref{ss:pert_fiedlds}. So that, expanding \Eref{eq:fields_decomposition}$^2$ by means of \Eref{eq:eshelby} gives the local strains inside $r$-th inclusion in the form
\begin{equation}
  \tenss{\strain}_r = \tenss{\strain}^0 + \tensf{S}_r \dcontr \tenss{\strain}^\tau_r \stackrel{\mathrm{\Eref{eq:eq_incl_meth_step_5}}}{=} \tenss{\strain}^0 + \tensf{S}_r \dcontr \B_r \dcontr \tenss{\strain}^0 = \tensf{A}_r^\mathrm{dil} \dcontr \tenss{\strain}^0
  \label{eq:dilute_1}
\end{equation}
where
\begin{equation}
  \tensf{A}_r^\mathrm{dil} = (\identity + \tensf{S}_r \dcontr \B_r)
  \label{eq:Ar_dilute}
\end{equation}
By analogy, considering $\tenss{\Stress}$ to approach $\tenss{\stress}^0$ and taking into account \Eref{eq:stress_hom} and \Eref{eq:eps_r}$^2$ gives
\begin{equation}
  \tensf{B}_r^\mathrm{dil} = [\identity + \B_r\dcontr(\tensf{S}_r - \identity)]
  \label{eq:Br_dilute}
\end{equation}

\subsubsection{Mori-Tanaka approximation}

The Mori-Tanaka approximation to concentration factors falls into the class of the so called mean-field theory methods. Namely, the inclusion interactions are accounted for by making use of the assumption that each inclusion is embedded separately in a large volume of a matrix which is subjected to a uniform remote stress or strain equal to as yet unknown averages~\cite{Mori1973571}.
In particular, the aim is to arrive at concentration factors $\tensf{A}_r^\mathrm{MT},~\tensf{B}_r^\mathrm{MT}$ as functions of the polarization tensors which are equal to dilute concentration factors from \Eref{eq:Ar_dilute} and \Eref{eq:Br_dilute}. Thus, the strain and stress in the $r$-th inclusion, respectively, reads as
\begin{equation}
  \tenss{\strain}_r = \tensf{A}_r^\mathrm{dil}\tenss{\strain}_0,\quad
  \tenss{\stress}_r = \tensf{B}_r^\mathrm{dil}\tenss{\stress}_0
  \label{eq:MT_1}
\end{equation}
From the strains averaged over the entire spectrum of $n$ inclusions plus that in the matrix phase, one can deduce, see e.g.~\cite{bohm1998short},
\begin{equation}
 \langle\tenss{\strain}\rangle = \bigg(c_0\identity + \sum_{r=1}^n c_r\tensf{A}_r^\mathrm{dil}\bigg)\tenss{\strain}_0
 \Rightarrow
 \tenss{\strain}_0 = \bigg(c_0\identity + \sum_{r=1}^n c_r\tensf{A}_r^\mathrm{dil}\bigg)^{-1} \langle\tenss{\strain}\rangle
  \label{eq:MT_2}
\end{equation}
Introducing~\Eref{eq:MT_2}$^2$ into~\Eref{eq:MT_1}$^1$, we arrive at
\begin{equation}
  \tenss{\strain}_r = \tensf{A}_r^\mathrm{dil}\bigg(c_0\identity + \sum_{r=1}^n c_r\tensf{A}_r^\mathrm{dil}\bigg)^{-1} \langle\tenss{\strain}\rangle = \tensf{A}_r^\mathrm{MT} \langle\tenss{\strain}\rangle = \tensf{A}_r^\mathrm{MT} \tenss{\Strain}
  \label{eq:MT_4}
\end{equation}
such that entering back to \Eref{eq:basic_averaging_rule_compl}$^1$ gives effective stiffness moduli stored in $\stiffmacr$. By analogy, starting the above analysis from \Eref{eq:MT_1}$^2$ gives the Mori-Tanaka approximation to stress concentration factor $\tensf{B}_r^\mathrm{MT}$ in the form
\begin{equation}
  \tensf{B}_r^\mathrm{MT} = \tensf{B}_r^\mathrm{dil}\bigg(c_0\identity + \sum_{r=1}^n c_r\tensf{B}_r^\mathrm{dil}\bigg)^{-1}
  \label{eq:MT_5}
\end{equation}
yielding effective compliance moduli by making use of \Eref{eq:basic_averaging_rule_compl}$^2$.

\subsubsection{Self-Consistent approximation}

Interactions among $r = 1,\dots,n$ phases are accounted for by assuming that each phase is an inclusion placed in the homogeneous medium of yet unknown overall properties of the aggregate of $n\backslash r$ remaining inclusions. It thus falls into the class of the so called effective medium methods. The Self-Consistent method is known to overestimate the interaction influence~\cite{gueguen1997microstructures}, which makes it specifically tailored for particulate media where a matrix phase, usually formed by fine particles, can not be clearly distinguished. Contrary to the approximations presented above, the Self-Consistent method results in implicit formulas~\cite{Hill1965213}.
Starting from the dilute approximation one can write
\begin{equation}
  \tensf{A}_r^\mathrm{SC} = (\identity + \tensf{S}_r^\mathrm{SC} \dcontr \B_r^\mathrm{SC}),\quad
  \tensf{B}_r^\mathrm{SC} = [\identity + \B_r^\mathrm{SC}\dcontr(\tensf{S}_r^\mathrm{SC} - \identity)]
  \label{eq:SC_Ar_Br}
\end{equation}
where the superscript $\bullet^\mathrm{SC}$ denotes explicit dependence of a quantity on material moduli coming from the Self-consistent approximation. In other words, stiffness moduli entering the formulas for $\B$, \Eref{eq:remote_transf_strain_oper} and $\tensf{S}$, see e.g.~\cite{mura1987micromechanics}, are functions of $\tensf{C}^\mathrm{CS}$ by substitution for $\tensf{C}^0$, notice especially \ErefsIII{eq:stiff_decomposition}{eq:remote_transf_strain_oper}.

Note, that the so called Cai-Horii approximation is obtained after the first iteration of the Self-consistent scheme~\cite{cai1993constitutive} where the quantities on the right-hand sides of both terms in \Eref{eq:SC_Ar_Br} are functions of the properties coming out the dilute approximations in \Erefs{eq:Ar_dilute}{eq:Br_dilute}.

\subsubsection{Differential scheme}

The differential scheme also falls into the family of effective medium methods. Contrary to the Self-Consistent approximation, this method builds the effective medium by incrementally adding inclusions to the matrix of effective properties obtained in previous steps. For instance, in the first step, \mumech{} adds the first inclusion to the virgin matrix of stiffness $\tensf{\stiff}_0$. In the next step, it adds another inclusion from the list to the matrix of effective properties obtained from the dilute approximation to the first step problem, and so on. It is clear that the previously homogenized matrix is not isotropic anymore unless the first inclusion was of the circular or spherical shape. That is why, \mumech{} performs numerical integration of elliptic potentials entering \Eref{eq:eshtensor_ext} for the Eshelby tensor $\tensf{\eshtens}$, see e.g. \cite{gavazzi1990numerical,vorel2013homogenization} for more details.

\section{Implementation}\label{sec:implementation}
%
Recall that the \mumech{} library was primarily designed as a module of finite element packages. Its main goal is the evaluation and post-processing of macro-field perturbations, which may take over the role of microstructure-informed enrichments for partition of unity strategies. So far, the code is furnished with analytical solutions to two and three dimensional problems with inclusions of ellipsoidal shapes, such as an ellipse or a circle in two dimensions and an ellipsoid, sphere, oblate spheroid, prolate spheroid, penny, flat ellipsoid, cylinder, and elliptic cylinder in three dimensions. The 3D and 2D inputs can not be mixed as the library runs in either of the modes at a single instance.
The functions are tuned in a way that inclusion of an arbitrary shape can be treated as a general ellipsoid with one or more degenerated semi-axes, e.g. a cylindrical fiber can be modeled as the ellipsoid with excessive $a_1$ semi-axis. By analogy, 2D plane strain conditions can be simulated as 3D cylinders with very long semi-axes parallel to global $z$ coordinate. However, in this case, the solution losses from its computational efficiency and specific shapes should be preferred instead the degenerated ones, namely in situations when dealing with large numbers of inclusions (in orders of millions).
Therefore, the inclusions defined as general ellipses/ellipsoids are automatically assigned relevant shapes according to the particular semi-axes dimensions by default.

A longer term ambition of the \mumech{} developers is to cover a maximum topics tackled by the micromechanics community. Therefore, the current release was also equipped with the classical homogenization techniques as discussed in \Sref{sec:background}. Moreover, to the best of our knowledge, the presented library is the only of its kind freely available at the time being.\todo{proverit  popripade posledni vetu presunout do introduction}

\subsection{Implementation scheme/Data flow structure}\label{sec:scheme}

\begin{figure}[!t]
  \centering
  \includegraphics[width=140mm]{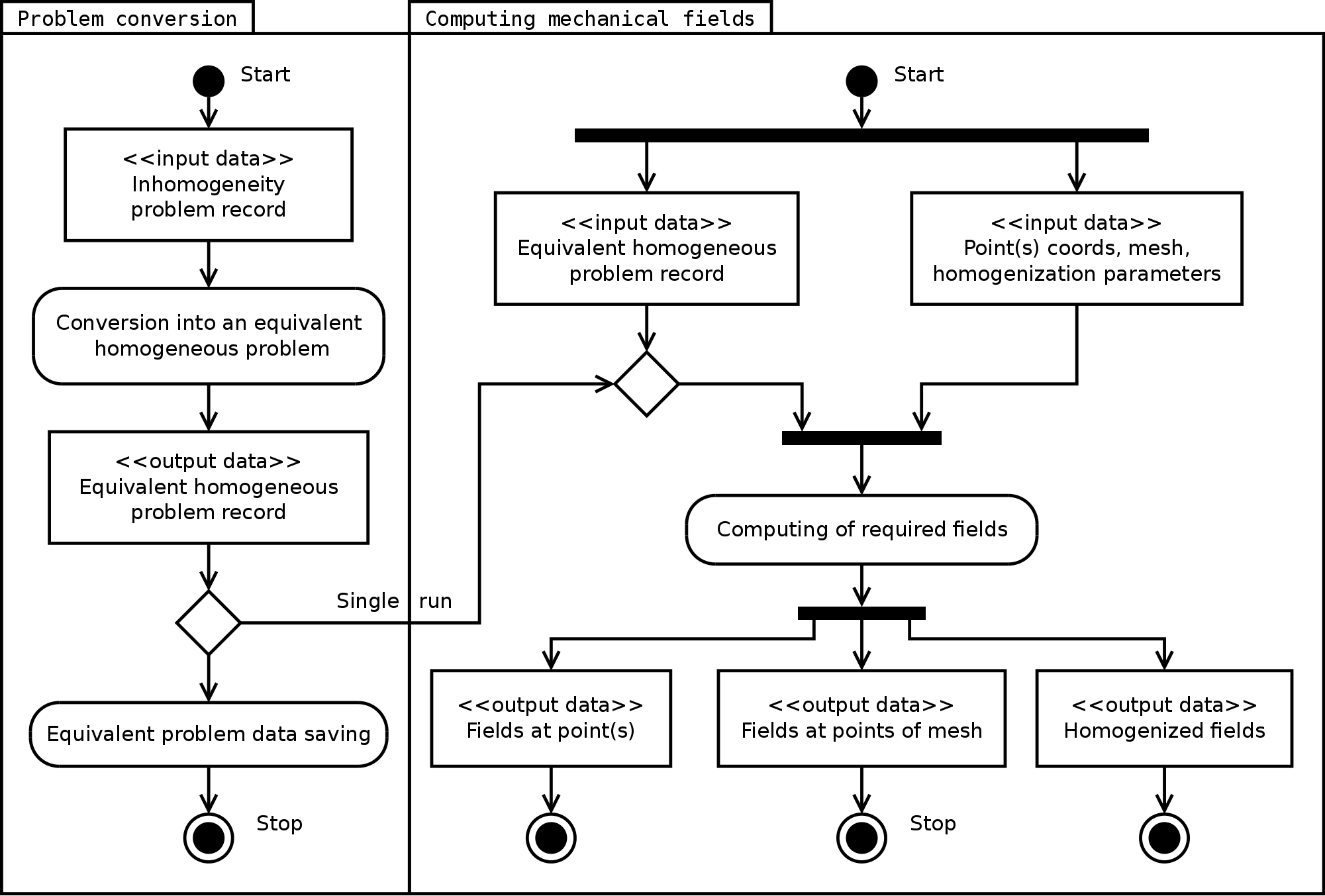}
  \caption{Implementation scheme.}
  \label{fig:scheme}
\end{figure}

The general structure of \mumech{} is briefly outlined in~\reffig{scheme}. Basically, it splits in two major tasks, conversion of inhomogeneity problem to equivalent inclusion problem and evaluation of mechanical fields eventually homogenization. The first step is usually most time-consuming, however \rev{it needs to be performed only once} and the appropriate data can be stored for subsequent analyzes over the same data, geometry and distribution of inhomogeneities to be exact\footnote{Note again that the primary purpose of \mumech{} is feeding FE packages with subscale data.}. The second branch of the algorithm can be called repeatedly to evaluate fields at different locations or to run different homogenization algorithms. However, in the single run cases, the code allows analyzes without saving and reading the auxiliary data.

Description of the inhomogeneity problem is required as the input for the first part of the algorithm. It consists of the geometry definitions (centroids, dimensions and rotation of semi-axes), material characteristics of the inhomogeneities and the matrix (Youngs modulus and Poisson's ratio), and definitions of imposed eigenstrains\footnote{Due to the induced pore pressure or thermal expansion for example.} and the remote strain tensors. The latter mentioned remote strains are handled as individual load cases and as such their number is arbitrary. The inhomogeneity inputs are converted into the equivalent problem by making use of Equivalent inclusion method and the self-compatibility procedure presented in~\Sref{sec:background}. The stored data, if required, are the equivalent transformation eigenstrains $\tenss{\strain}^\tau_r$ and local (say internal) Eshelby tensors $\tensf{\eshtens}_r$.

In the second part of the algorithm, mechanical fields at user-defined points, including those outside inclusions, are evaluated. Individual $\x$ coordinates can be entered one by one or in arbitrarily large sets, e.g. nodes or integration points of an FE mesh. Calculated fields may be postprocessed with the in-built post-processor and visualized with tools as Paraview, MayaVi, etc.\cite{Paraview,MayaVi}. Another in-built feature is the homogenization of calculated local fields by DIM introduced in~\Sref{sec:homogenization}. Optionally, users can disable the self-compatibility algorithm and the evaluation of local fields and use alternative micromechanical approaches discussed also therein.

\subsection{I/O data specification}\label{sec:IO}
%

%
%
\begin{table}[t!] \centering
  {\tableset{0}{6}
    \begin{tabular}{|m{135mm}|}
      \hline \vspace{0mm}
	     {\footnotesize
\begin{verbatim}
# vtk DataFile Version 3.0
3D - example, 2 inclusions
ASCII
DATASET UNSTRUCTURED_GRID
POINTS 2 float
-1.0   1.0  0.0
 2.0   0.0  0.0
POINT_DATA 2
VECTORS Semiaxes_dimensions float
1.0 1.0 1.0
1.0 0.7 0.4
VECTORS Euller_angles_deg float
0.0 0.0 0.0
35.0 0.0 0.0
SCALARS Youngs_modulus float 1
LOOKUP_TABLE default
5.5
2.4
SCALARS Poissons_ratio float 1
LOOKUP_TABLE default
0.3
0.3
TENSORS Imposed_eigenstrains float
0.0 0.0 0.0 0.0 0.0 0.0 0.0 0.0 0.0
0.0 0.0 0.0 0.0 0.0 0.0 0.0 0.0 0.0
FIELD unstructured_data 2
Matrix_record 1 2 float
1.0  0.4
Remote_strains 9 3 float
1.0 0.0 0.0 0.0 0.0 0.0 0.0 0.0 0.0
0.0 0.0 0.0 0.0 1.0 0.0 0.0 0.0 0.0
2.0 1.5 0.0 1.5 0.0 0.0 0.0 0.0 3.0
\end{verbatim}
	     }
	     \\[0mm] \hline
    \end{tabular}
  }
  \caption{Double inclusion problem input file in legacy VTK format.}
  \label{tab:VTKtab}
\end{table}

The library is designed in a way a user or a master program invokes the feedback by using a set of C++ functions. Despite, the I/O data flow between \mumech{} and a governing instance can be realized via parameters of the interface functions, the exchange by means of files is also possible as it proved to be more practical especially for large number of inclusions. In both cases, the data have a unified syntax. Symmetric tensors are handled in a non-reduced form and together with non-symmetric tensors, e.g. $\tensf{\eshtens}$, are stored in row-by-row vectors, in the so called Iliffe arrays. In 2D mode the input data can be reduced correspondingly, i.e. coordinates may have only two components, $2$-nd order tensors are of dimensions $2\times2$, etc. In particular, following data are handled by means of files, the inhomogeneity and equivalent homogeneous problem records, grids of point coordinates\footnote{it can be e.g. an FE triangulation} in which the mechanical fields are evaluated, and finally the sought fields themselves, see~\Fref{fig:scheme}. The ASCII Visualization Tool Kit (VTK) format in both legacy and XML variants has been chosen as the native file syntax~\cite{VTKpdf}, as it is human-readable and can be visualized directly in a modeler or free visualization tool-kits~\cite{Paraview,MayaVi}. Therefore, the data can be easily controlled at any stage of the software development, debugging, or most importantly, in a routine use.

An example of the legacy VTK file with a composite media description is shown in~\reftab{VTKtab}. It describes the 3D matrix with a spherical and an ellipsoidal inclusions loaded by three remote strains. The dimension of the problem is explicitly determined by the ``3D'' keyword at the beginning of the second line, which is originally reserved for comments. The number of inclusions and centroid coordinates are given in the data block following the keyword ``POINTS''. Dimensions and rotation of semi-axes, and material characteristics of each inclusion are listed in the section introduced by ``POINT\_DATA''. Finally, the data describing the infinite medium are specified in the section preceded by the ``FIELD'' keyword. In particular, these are Youngs modulus and Poisson's ratio of the matrix and the remote strain tensors.

\subsection{Interface functions}\label{sec:interfces}

The class {\code Problem} is the central element, better say a type, of the object-oriented source code and the vast majority of \mumech{} features is accessed through its public members. A representative implementation calling crucial functions of the inhomogeneity problem analysis is listed in \Tref{lis:mumech_example}.

\lstset{
	basicstyle=\ttfamily\small,
	captionpos=b,
	numbers=left,
	numbersep=8pt,
	xleftmargin=2em,
}

\begin{table}
\begin{lstlisting}[language=c++]
Problem *p = new Problem;
p->read_input_file("inhomogeneity.vtk");
p->input_data_initialize_and_check_consistency();
p->convert_to_equivalent_problem();
p->print_equivalent_problem("equivalent.vtk");
delete p;
p = new Problem;
p->read_input_file("equivalent.vtk");
p->input_data_initialize_and_check_consistency();
double coords[] = {0.5,0.0,0.0};
double **stress = AllocateArray2D(2,9);
p->giveFieldsOfPoint(NULL,NULL,stress,coords,'p',0,2);
p->printFieldsOnMeshVTK("results.vtk","mesh.vtk",'t',2,1);
p->print_visualization("visualization.vtk",5);
delete p;
DeleteArray2D(stress,2);
\end{lstlisting}
\caption{Example code of \mumech{} interface.}
\label{lis:mumech_example}
\end{table}

Line 2 is responsible for importing a complete problem description from the VTK file listed in~\reftab{VTKtab}. The data initialization and verification follows in line 3. Line 4 converts the inhomogeneity problem into the equivalent inclusion problem. The data for multiple use of the problem geometry are stored in line 5, if required. As demonstrated in lines 6-9, imports of both, the inhomogeneity and equivalent homogeneous problems work in the same fashion. Clearly, lines 5-9 or 6-9 can be omitted in the case of a single run.

The function {\code giveFieldsOfPoint} evaluates mechanical fields at a given point, lines 10-12. Displacement, strain and stress fields are returned by means of the first three parameters, respectively. Each of the parameters is a double pointer to the two-dimensional array. The first dimension is equal to the number of load cases, i.e. remote strains specified at the end of the input file, \reftab{VTKtab}. The second dimension equals the length of the vector, in the case of displacements, or row-by-row stored tensors when recalling strains or stresses. Passing {\code NULL} pointer indicates that the corresponding quantities will not be calculated.
The fourth parameter is a pointer to an array of the point coordinates. The next {\code char} parameter denotes the character of evaluated fields where 'p' stands for perturbations while 't' for their total counterparts, see~\Sref{sec:background}. Finally, the last two parameters determine the index of the first load case and number of load cases to be comprised in the analysis. In this particular case, a pair of perturbation stress tensors due to the first (0-th in C-like syntax) and second remote strain excitations are evaluated in line 12.
The data visualized in Paraview are shown in~\Fref{fig:3d_example_stress}b.

In line 13, function {\code printFieldsOnMeshVTK} reads the FE mesh from {\it mesh.vtk} file and evaluates total fields in element nodes and the third given remote strain. Then the mesh with results is stored in the file {\it results.vtk}. Finally, line 14 performs triangulation of inclusion surfaces which is written in {\it visualization.vtk} file. The data visualized in Paraview are shown in~\Fref{fig:3d_example_stress}a.

A detailed description of other interface functions can be found in tutorial~\cite{mumech} together with a number of ways how to control the analysis, e.g. functions for running different homogenization algorithms, the switch parameter between {\it full} and {\it optimal} version of either the self-compatibility procedure or evaluation of perturbation/total fields at different points, etc.

\subsection{Technology}\label{sec:Technology}

\mumech{} is a free open source software. It can be run, modified, and redistributed under the terms of the GNU Lesser General Public License as published by the Free Software Foundation; either version 2 of the License, or any later version~\cite{GNU:LGPL}. The project has been implemented in C++, as it is easier to maintain accessibility of its generic structure while it also enables robust low level optimization of time-consuming algorithms. Multi platform CMake~\cite{CMake} was chosen to configure and build source code properly on client machines. A complete source code and documentation generated by Doxygen~\cite{Doxygen} script can be found at \http{mumech.cz}, together with a number of examples and input files by means of which we perform the compilation of an executable file and testing. All interface functions and examples are documented in a tutorial also available at the project website.

\section{Numerical examples and performance}\label{sec:examples}

\begin{figure}[!ht]
  \centering
   \begin{tabular}{cc}
    \includegraphics[height=55mm]{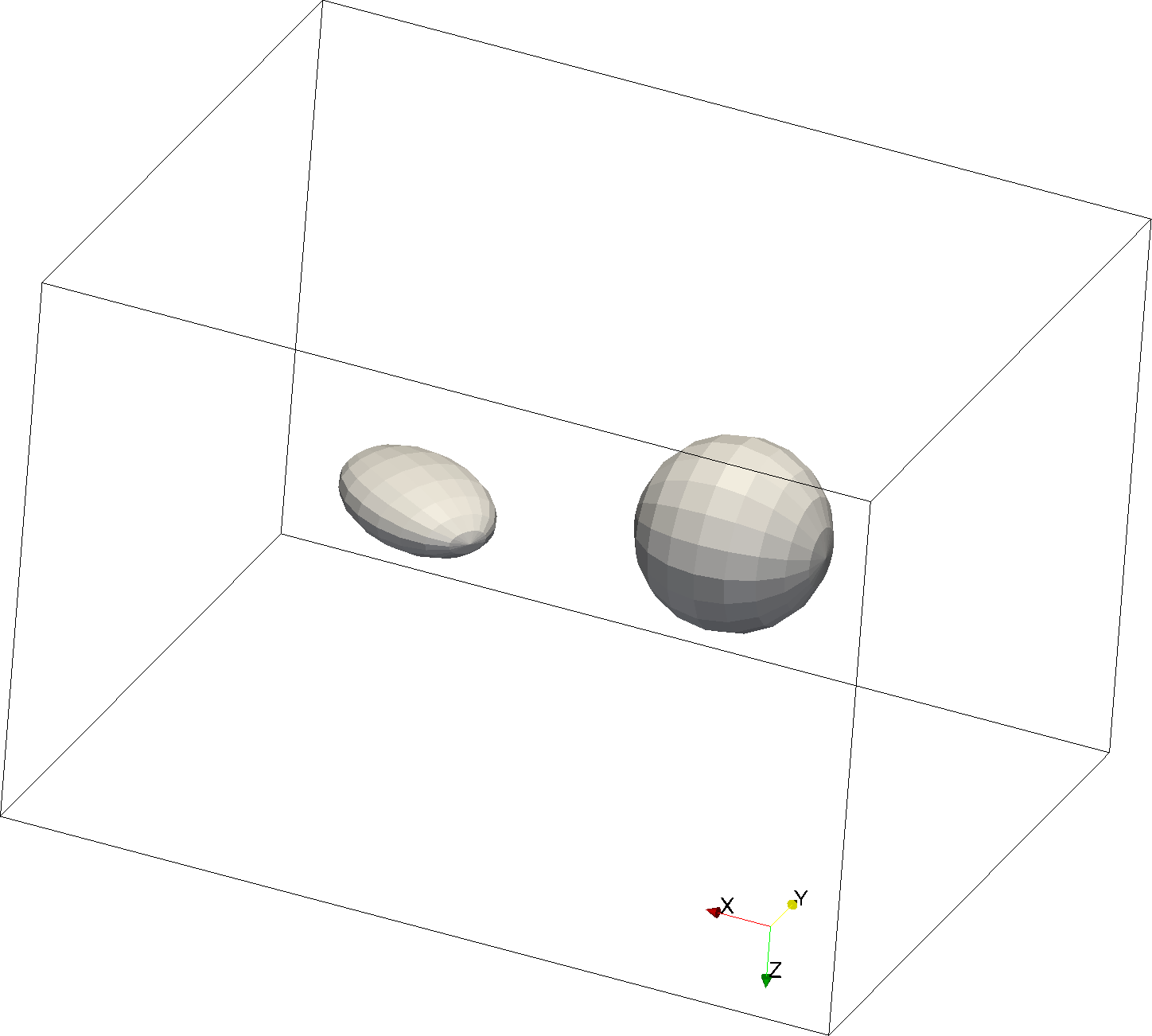}&
    \includegraphics[height=55mm]{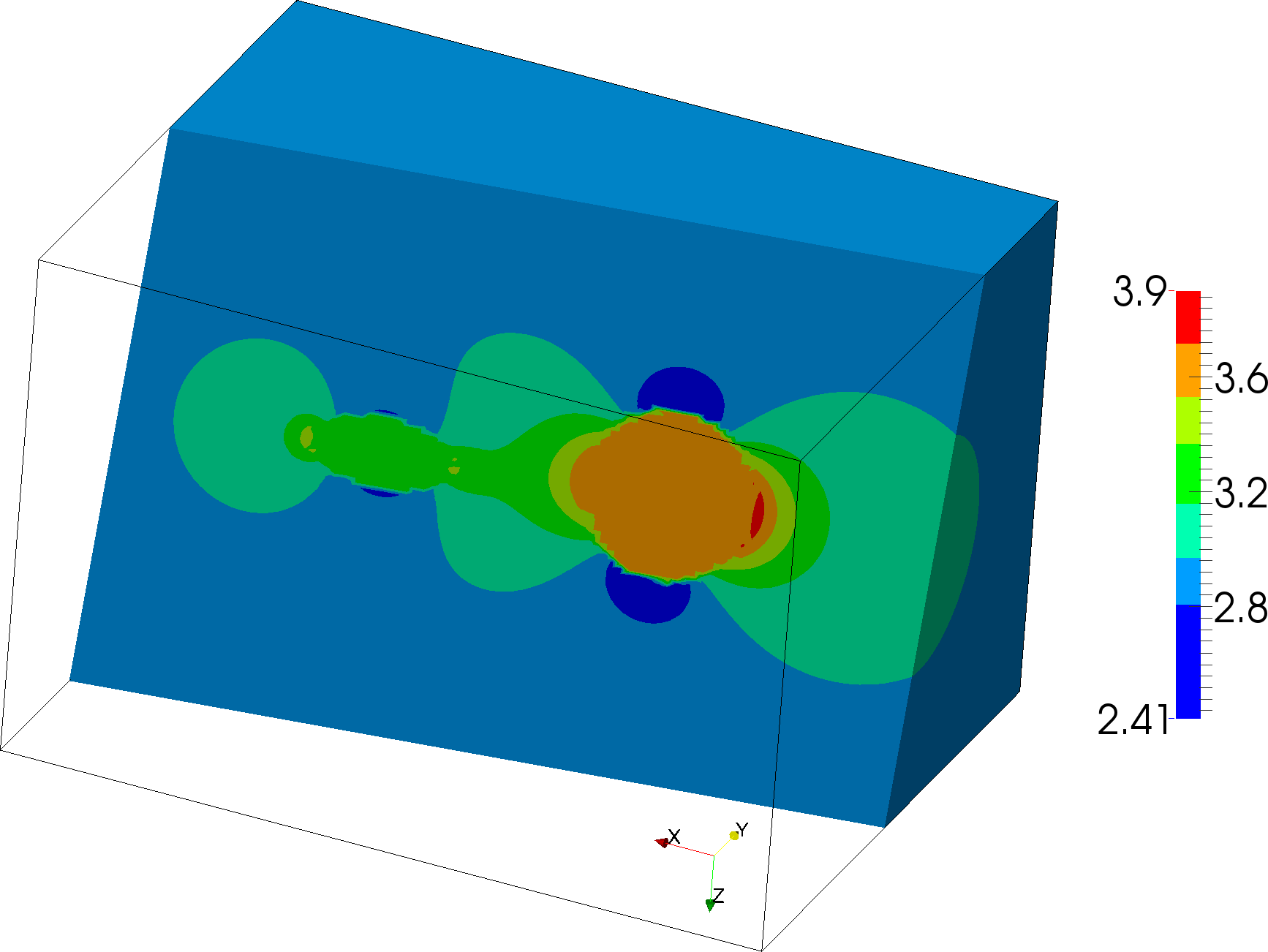}\\
     (a) &(b)
   \end{tabular}
  \caption{Example of 3D double inclusion problem, a) benchmark geometry, b) patterns of $\stress_{11}$.}
  \label{fig:3d_example_stress}
\end{figure}

\begin{figure}[t]
  \begin{center}
    \begin{tabular}{cc}
    \resizebox{.5\textwidth}{!}{ 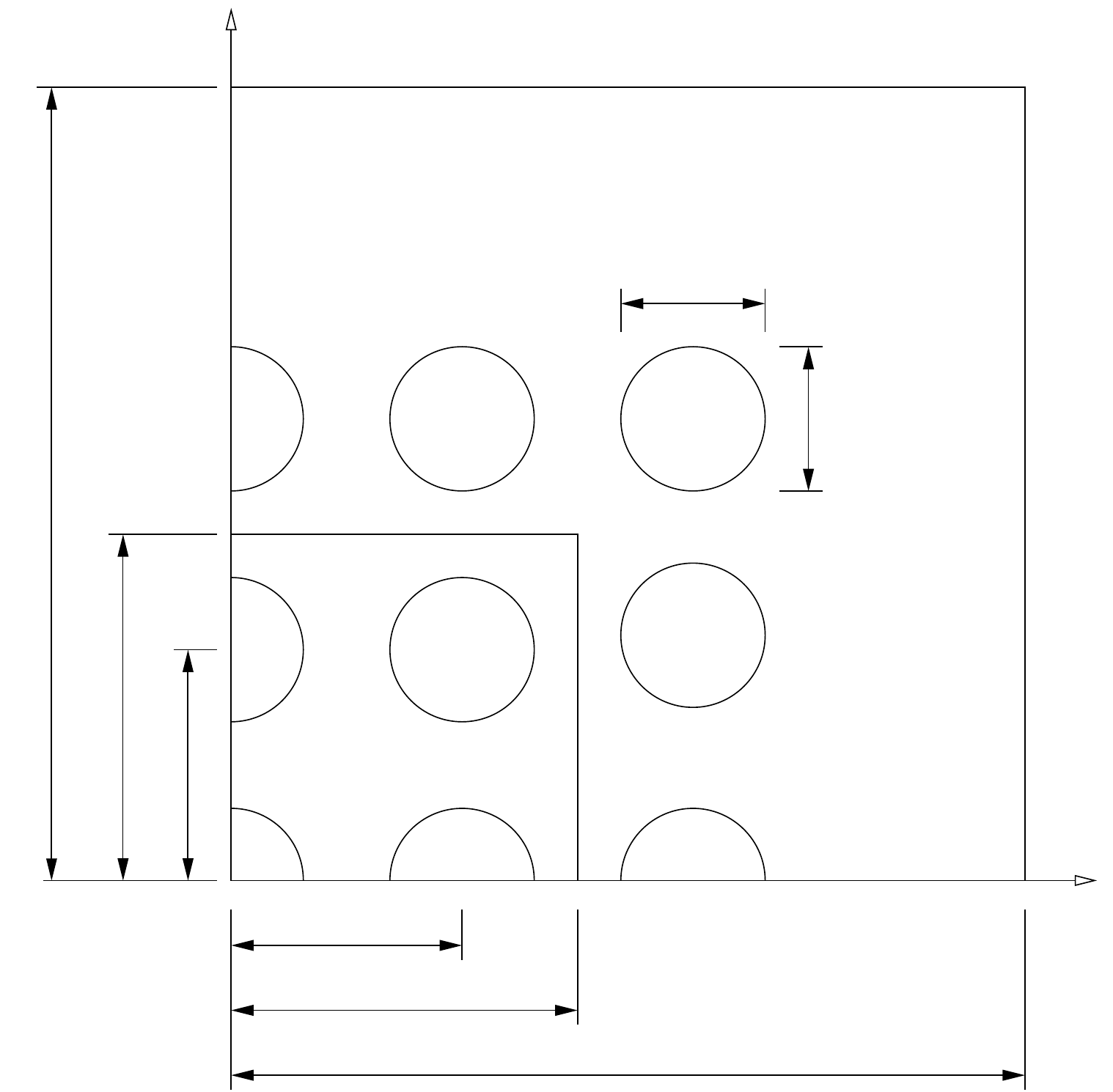 }&
    \includegraphics[height=70mm]{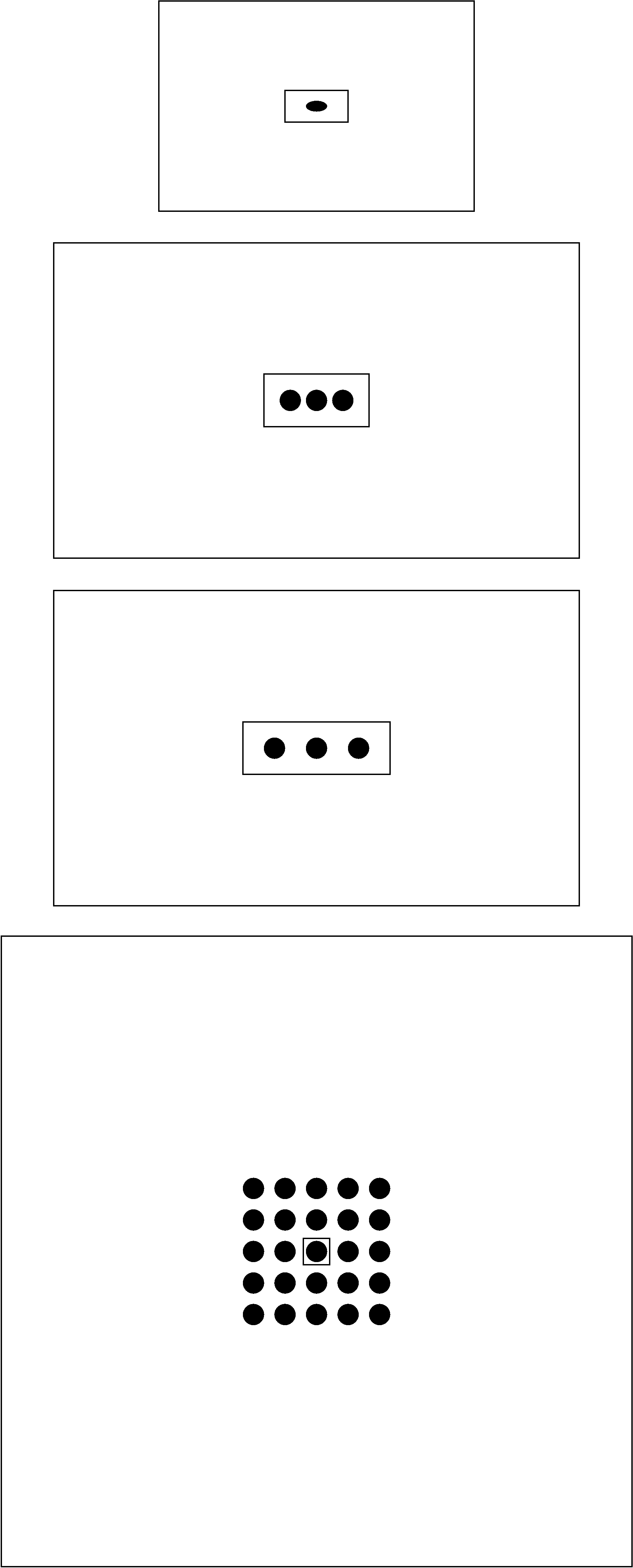}\\
    (a) &(b)\\
    \end{tabular}
    \caption{Geometry of 2D benchmarks, a) parametric setting of tested benchmarks, b) particular geometry of single, triple - narrow gaps, triple - wider gaps, and multiple inclusion test.}
    \label{fig:grid}
  \end{center}
\end{figure}

\begin{figure}[!ht]
  \centering
  \begin{tabular}{cc}
  \imagetop{\includegraphics[height=44mm]{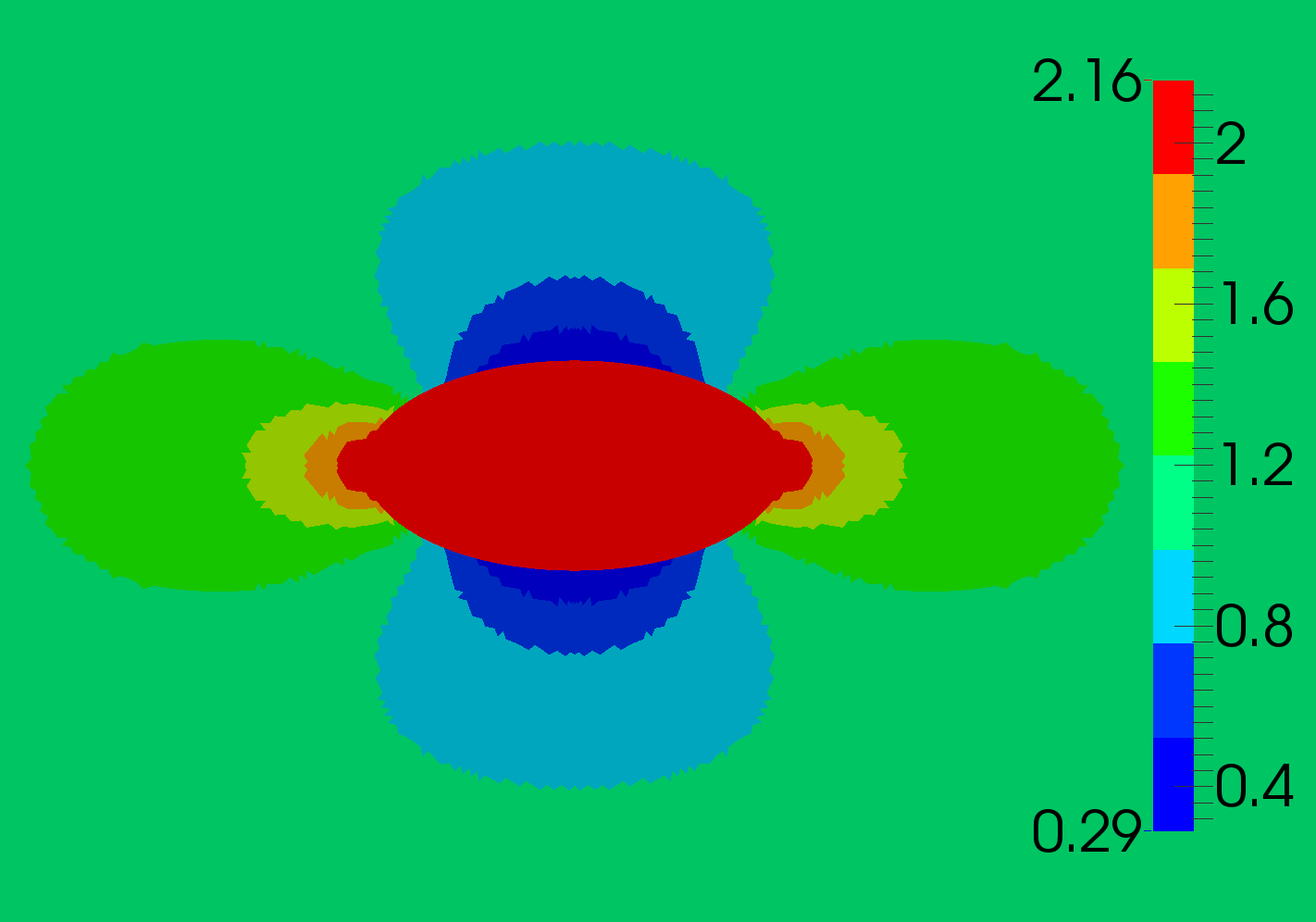}}&
  \imagetop{\includegraphics[height=51mm]{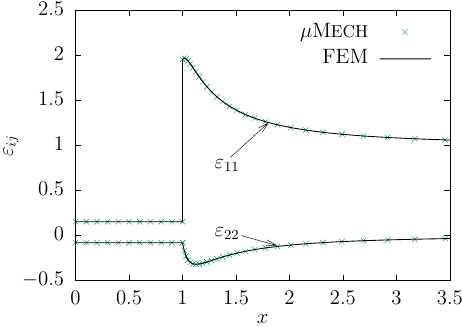}}\\
  (a) &(b)
  \end{tabular}
  \caption{Single inclusion test, a) patterns of $\stress_{11}$ calculated by \mumech{}, b) total strain components along $x$-axis.}
  \label{fig:graf_1I}
\end{figure}

\begin{figure}[!ht]
  \centering
  \begin{tabular}{cc}
    \includegraphics[height=50mm]{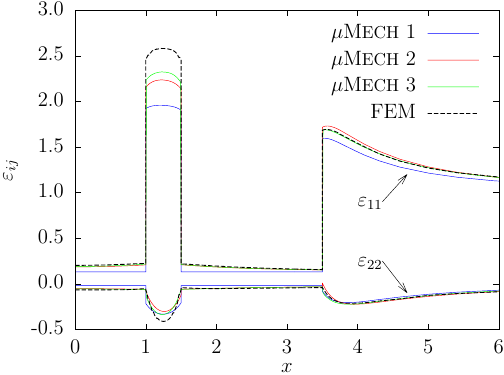}& \includegraphics[height=50mm]{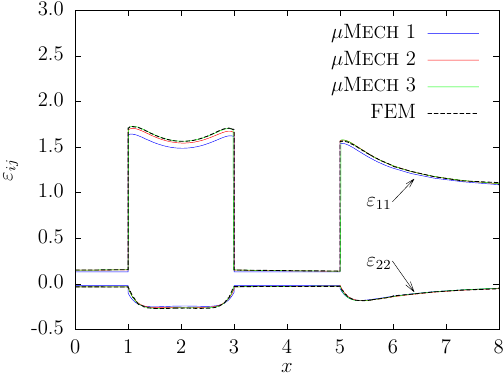}      \\
    (a) &(b)                                                                                                  \\
    \includegraphics[height=50mm]{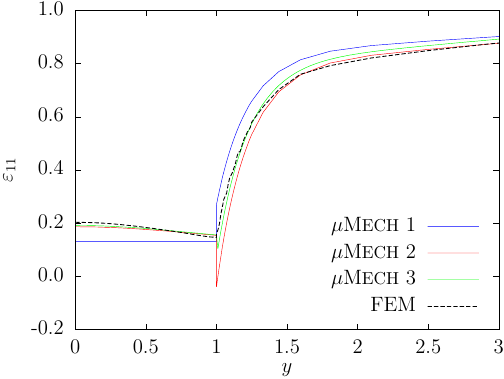}& \includegraphics[height=50mm]{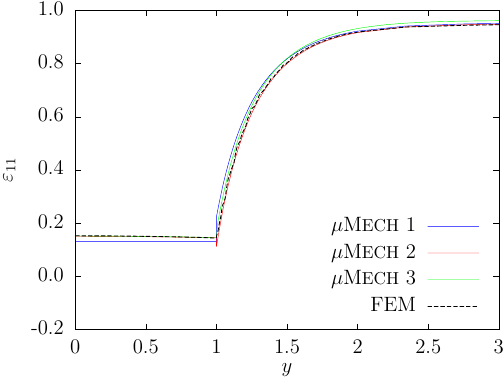}  \\
    (c) &(d)                                                                                                  \\
    \includegraphics[height=50mm]{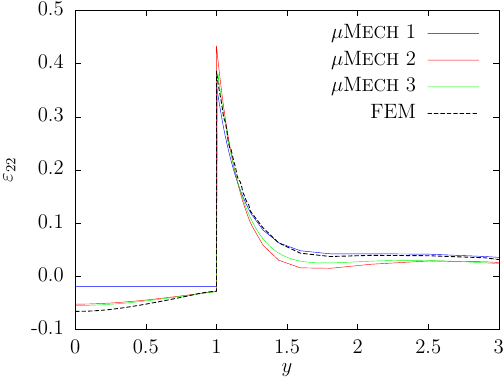}& \includegraphics[height=50mm]{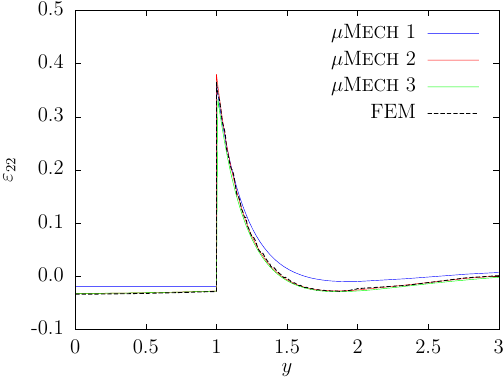}  \\
    (e) &(f)
  \end{tabular}
  \caption{Triple inclusion tests, a,b) total strain components along $x$-axis for inclusions with narrow and wider gaps, respectively, c,d) total strain components along $y$-axis for inclusions with narrow and wider gaps, respectively.}
  \label{fig:graf_3I}
\end{figure}

\begin{figure}[!ht]
  \centering
  \begin{tabular}{c}
    \includegraphics[height=50mm]{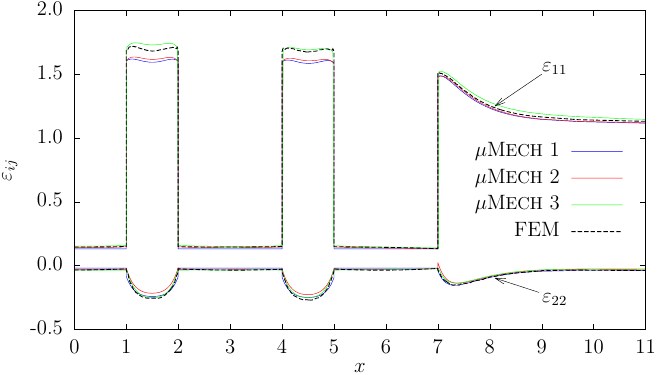}\\
    (a)\\
    \includegraphics[height=50mm]{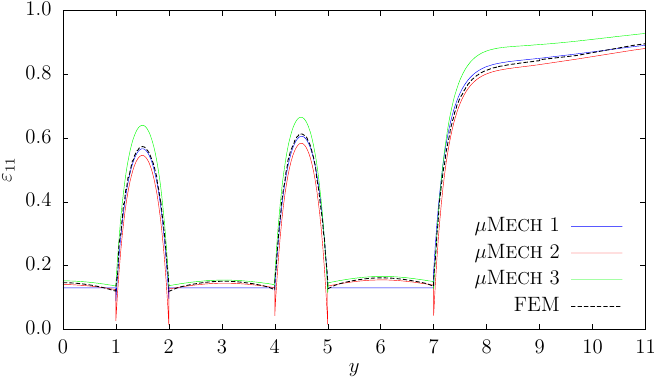}\\
    (b)\\
    \includegraphics[height=50mm]{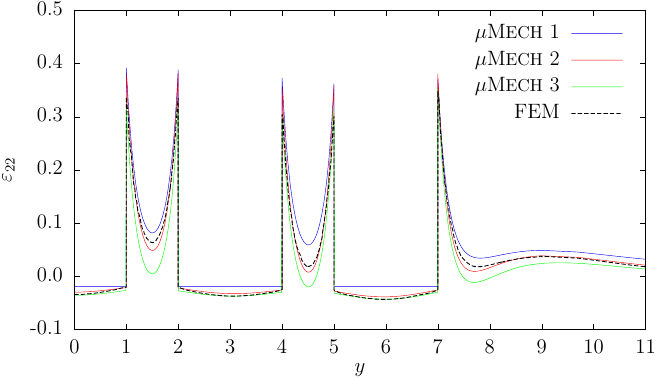}\\
    (c)
  \end{tabular}
  \caption{Multiple inclusion test, a) total strain components along $x$-axis, b) total strain components along $y$-axis.}
  \label{fig:graf_5I}
\end{figure}

The capabilities of the \mumech{} library are briefly demonstrated through a 3D double inclusion task and a series of 2D multiple inclusion examples under plane strain conditions. The 3D analysis is composed of a pair of inclusions, one ellipsoid and sphere. The geometry, topology and material parameters together with the prescribed strain excitation are specified in \Tref{tab:VTKtab}. In addition, the geometry triangulated by \mumech{} and visualized in Paraview is shown in \Fref{fig:3d_example_stress}a while the distribution of axial stress $\stress_{11}$ is plotted in \Fref{fig:3d_example_stress}b.
In order to discuss the quality of solutions by \mumech{} we have compared the 2D analyzes with reference solutions by FEM. The tests were performed in the 2D setting for the better visualization purposes, however we have executed the same calculations by means of the 3D implementation with degenerated semi-axis and arrived at exactly the same results. The first task is the single elliptic inclusion problem. The second and third tasks are the triple inclusion tests with centroids of circular inclusions aligned in $x$ direction. The two tasks differ in the mutual distances among the inclusions. The last test comprises 25 circular inclusions distributed in a regular grid of $5\times 5$ points in $x-y$ plane and representing inclusion centroids. The geometry of all four tests is given by the parameters in \Tref{tab:task_geom_params} whose meaning is evident from \Fref{fig:grid}.
The material parameters were set to $E_r = 10.0, \nu_r = 0.3, E_0 = 1.0, \nu_0 = 0.2$ for all the analyzes. Finally, the remote strain excitation imposed to the infinite matrix was such that $\tenss{\strain}^0_{11} = 1.0$ while the other components vanished. Note, in the case of FE comparative analyzes, appropriate the remote strains were imposed by means of the boundary displacements $u_i^\mathrm{fem}$ applied on $\boundary_0$. The particular magnitude of $u_i^\mathrm{fem}$ is evident from \Fref{fig:grid}a.
\begin{table}[h] \centering
  {\tableset{1.05}{6}
    \begin{tabular}{|c|c|c|c|c|c|c|c|c|}
      \hline
      No. inclusions   & $a_1$ & $a_2$ &  $d_1^I$ & $d_2^I$ &  $d_1^G$ & $d_2^G$ & $d_1^O$ & $d_2^O$ \\      \hline
      $1 \times 1$      & 1.0   & 0.5   &  -      &  -      &  3.0     &  1.5    & 15.0   & 10.0    \\      \hline
      $3 \times 1$ (narrow gaps)      & 1.0   & 1.0   &  2.5    &  -      &  5.0     &  2.5    & 25.0   & 15.0    \\      \hline
      $3 \times 1$ (wider gaps)      & 1.0   & 1.0   &  4.0    &  -      &  7.0     &  2.5    & 25.0   & 15.0    \\      \hline
      $5 \times 5$      & 1.0   & 1.0   &  3.0    &  3.0    &  1.5     &  1.5    & 30.0   & 30.0    \\      \hline
    \end{tabular}
  }
  \caption{Geometrical and topological parameters of four 2D tasks performed.}
  \label{tab:task_geom_params}
\end{table}
The qualitative comparison of the three types of solutions, (i) a solution without performing self-compatibility algorithm (labeled as \mumech{}~1 in the sequel), (ii) a solution including the adjustment by means of the self-compatibility algorithm and the non-constant approximation to internal fields as proposed in \Pref{sec:approx_concentrations}, and (iii) the previous solution \mumech{}~2 enhanced by the approximation to external fields calculated by means of linear transformation eigenstrains (labeled as \mumech{}~3). The distribution of $\stress_{11}$ for the single inclusion problem calculated by \mumech{} is shown in \Fref{fig:graf_1I}a. The comparison with the FEM solution in terms of total strain components in $x$--axis direction is evident from \Fref{fig:graf_1I}b.
The series of figures with individual strain components compared with respect to FE solutions for remaining tasks are displayed in \Fref{fig:graf_3I} and \Fref{fig:graf_5I}. Note namely the obvious local convergence of individual \mumech{} methods 1--3 to the reference solution.
In the case of multiple inclusions, the mechanical fields within individual inclusions are not uniformly distributed as a result of their mutual interaction. There is an evident difference for strains taking place in the matrix, namely for inclusions positioned close to each other,~\Fref{fig:graf_3I}a and \Fref{fig:graf_5I}. However, the mutual interactions quickly disappear with increasing spacing as shown in~\reffig{graf_3I}b. An interesting behavior can be observed in \Fref{fig:graf_5I}b,c from which it is obvious that \mumech{} 3 method looses in $y$--direction with respect to its 1--2 counterparts. The reason is the low polynomial order, linear to be exact, of $\tenss{\strain}^\tau(\x)$ as indicated by analyzes with quadratic eigenstrains. A detailed justification of this hypothesis, however, is let for future work as the current implementation of the solution with quadratic eigenstrains is not furnished with analytical derivatives of elliptic potentials and the numerical differentiation is unstable enough to disable reliable testing.
Looking carefully at \Fref{fig:graf_3I}c and \Fref{fig:graf_5I}b one can observe Gibbs-like phenomenon at the inclusion interfaces related to the fact the solution to external fields is constructed as the sum of individual contributions from all $n$ inclusions entering the analysis, for details see \Sref{sec:background}. Clearly, this is nonphysical, though inevitable behavior that must be accepted when using current version of the \mumech{} library.

The quality of the \mumech{} solutions was further quantified in an average sense by means of the normalized error defined as
\begin{equation}
  \mathrm{err} = \frac{\norm{\tenss{e}}}{\norm{\tenss{e}^\mathrm{tot}}} \times 100\%
  \label{eq:err}
\end{equation}
where $\norm{\tenss{e}} = \sqrt{\int_{\domain_\Box} \tenss{e}_\strain \dcontr \tensf{C} \dcontr \tenss{e}_\strain \de{\domain}}$, $\norm{\tenss{e}^\mathrm{tot}} = \sqrt{\int_{\domain_\Box} \tenss{\strain} \dcontr \tensf{C} \dcontr \tenss{\strain}\de{\domain}}$, and $\tenss{e}_\strain = \tenss{\strain} - \tenss{\strain}^\mathrm{fem}$.
\begin{table}[h] \centering
  {\tableset{1.05}{6}
    \begin{tabular}{|c|c|c|c|}
      \hline
       Evaluation          &\multicolumn{3}{c|}{No. inclusions}\\      \cline{2-4}
       method              &  $3 \times 1$ (narrow gaps)  &  $3 \times 1$ (wider gaps)  &  $5 \times 5$ \\      \hline
      \mumech{} 1      &  17.1                        &   3.9                       &  4.5          \\      \hline
      \mumech{} 2      &  7.6                         &   1.7                       &  3.2          \\      \hline
      \mumech{} 3      &  6.9                         &   1.4                       &  2.6          \\      \hline
    \end{tabular}
  }
  \caption{Normalized errors according to \Eref{eq:err} and measured in \%.}
  \label{tab:norms}
\end{table}
The resulting values for the triplet of methods are listed in \Tref{tab:norms}. The results clearly show the superiority of the \mumech{} 3 method over the remaining two.
\begin{table}[h] \centering
  {\tableset{1.05}{6}
    \begin{tabular}{|l|c|c|c|c|c|}
      \hline
      Computation         & \multicolumn{3}{c|}{Stiffness tensor moduli}                     & \multicolumn{2}{c|}{Isotropic moduli}  \\\cline{2-6}
      \ctab{|c|}{scheme}  & $C_{1111} = C_{2222}$ & $C_{1112} = C_{2212}$ & $C_{1212}$ & $E$      &$\nu$                   \\\hline\hline
      Self-Consistent                  &  3.3461              &  1.0712            &  1.1374    &  2.82  &  0.24            \\
      Diff. Scheme              &  2.2149              &  0.6456            &  0.9093	   &  1.92  &  0.22            \\
      Mori-Tanaka         &  2.6811              &  0.8005            &  0.9402    &  2.31  &  0.22            \\
      Dilute              &  1.9309              &  0.5357            &  0.6976    &  1.69  &  0.21            \\\hline
      DIM $1\times1$      &  2.6833              &  0.7860            &  0.9264    &  2.31  &  0.22             \\
      DIM $3\times3$      &  2.8323              &  0.6265            &  0.8593    &  2.51  &  0.20             \\
      DIM $5\times5$      &  2.8417              &  0.6233            &  0.8541    &  2.52  &  0.20             \\
      DIM $7\times7$      &  2.8411              &  0.6261            &  0.8529    &  2.52  &  0.20             \\
      DIM $9\times9$      &  2.8406              &  0.6274            &  0.8525    &  2.52  &  0.20             \\\hline
      FEM                 &  2.8883              &  0.6531            &  0.8615    &  2.55  &  0.20             \\\hline
    \end{tabular}
  }
  \caption{Homogenized stiffness moduli for monodisperse with narrow gaps of $c_r = 0.5$. DIM results were obtained by means of \mumech{} 3 method. Isotropic moduli in last two columns were derived from eigenvalue analysis of $\stiffmacr$ as reported in \cite{dovskavr2016jigsaw}.}
  \label{tab:ncParams-verse}
\end{table}
The last analyzes performed cover the testing of homogenization approaches, namely that based on the direct integration -- DIM, and classical micromechanical schemes. Note, that in the case of DIM we took the integration domain $\domain_\Box$ as indicated in \Fref{fig:grid}a. Results for two different volume fractions $c_r$, proportional to the gaps among the inclusions as parametrized in \Tref{tab:task_geom_params}, are listed in \Tref{tab:ncParams-verse} and \Tref{tab:ncParams-verse_2}. The fit among all schemes is remarkable but the Self-Consistent scheme which is known to overestimate the moduli for lower volume fractions of stiff inclusions. Moreover, it appears that for both configurations, either narrow or wider gaps, $3\times 3$ inclusions adjacent to that inside $\domain_\Box$ is far sufficient for very accurate results.
\begin{table}[h] \centering
  {\tableset{1.05}{6}
    \begin{tabular}{|l|c|c|c|c|c|}
      \hline
      Computation         & \multicolumn{3}{c|}{Stiffness tensor moduli}                     & \multicolumn{2}{c|}{Isotropic moduli}  \\\cline{2-6}
      \ctab{|c|}{scheme}  & $C_{1111} = C_{2222}$ & $C_{1112} = C_{2212}$ & $C_{1212}$ &  $E$      &$\nu$                    \\\hline\hline
      Self-Consistent                  &  1.8511              &  0.5235            &  0.6639    & 1.62  &  0.22                 \\
      Diff. scheme              &  1.6587              &  0.4556            &  0.6630    & 1.46  &  0.21                 \\
      Mori-Tanaka         &  1.7417              &  0.4808            &  0.6304    & 1.53  &  0.21                 \\
      Dilute              &  1.5722              &  0.4228            &  0.5746    & 1.39  &  0.21                 \\\hline
      DIM $1\times1$      &  1.7524              &  0.4774            &  0.6272    & 1.54  &  0.21                 \\
      DIM $3\times3$      &  1.7889              &  0.4400            &  0.6046    & 1.59  &  0.20                 \\
      DIM $5\times5$      &  1.7904              &  0.4391            &  0.6036    & 1.59  &  0.20                 \\
      DIM $7\times7$      &  1.7905              &  0.4392            &  0.6033    & 1.59  &  0.20                 \\
      DIM $9\times9$      &  1.7905              &  0.4392            &  0.6032    & 1.59  &  0.20                 \\\hline
      FEM                 &  1.7854              &  0.4388            &  0.6017    & 1.58  &  0.20                 \\\hline
    \end{tabular}
  }
  \caption{Homogenized stiffness moduli for monodisperse with narrow gaps of $c_r = 0.35$. DIM results were obtained by means of \mumech{} 3 method. Isotropic moduli in last two columns were derived from eigenvalue analysis of $\stiffmacr$ as reported in \cite{dovskavr2016jigsaw}.}
  \label{tab:ncParams-verse_2}
\end{table}

\section{Conclusions}\label{sec:conclusions}

In the present paper we discussed a new and, to the best of our knowledge, the only freely available library of solutions to micromechanical problems based on Eshelby's seminal work~\cite{eshelby1957determination} and its subsequent extensive elaboration in classical textbooks as e.g.~\cite{mura1987micromechanics}. Contrary to what is meant as a standard in classical micromechanics, the implemented strategies aim at the evaluation of perturbation or total local fields inside and outside ellipsoidal inclusions. The code also covers the solution to multiple inclusion problems by means of the so called self-compatibility algorithm. This strategy benefits from the solution to the inclusion problem with polynomial eigenstrains. As this is the crucial part of the code we will keep improving it in the future, possibly with the help of new members of the emerging developers community motivated also by means of the present paper. Besides, the library is furnished with classical homogenization theories such as Mori-Tanaka, Self-Consistent method etc. On the basis of the above comments, let us stress current features of the \mumech{} library and a few proposals for further development as follows.

\noindent
Implemented features:
\begin{itemize}
\item solutions to internal and external fields in two and three dimensions,
\item an approximate solution to the multiple inhomogeneity/inclusion problem by means of the self-compatibility algorithm,
\item the solution to the equivalent inclusion problem with polynomial stress free transformation eigenstrains,
\item a powerful I/O interface based on the VTK standard,
\item various homogenization schemes as Mori-Tanaka, Self-consistent method, dilute approximation, direct integration, and the differential scheme.
\end{itemize}
Future development will focus on:
\begin{itemize}
\item a Galerkin-like approximation to the multiple inclusion problem with Eshelby functions at heart,
\item a direct link between \mumech{} core implementation and a F\# class for polynomial eigenstrain based solutions,
\item a parallelization of the solution to the multiple inclusion problem,
\item a special care of the Gibbs-like phenomenon taking place at the inclusion interfaces.
\end{itemize}

\section*{Acknowledgements}

The authors gratefully acknowledge the endowment of the Czech Science Foundation under the grant no. 13-22230S. We also thank Martin Do\v{s}k\'{a}\v{r} of CTU in Prague for careful reading of the manuscript and valuable comments on its scientific exposition.

\section*{References}

\end{document}